\documentclass[12pt, preprint]{aastex}

\usepackage{psfig, epsfig}
\usepackage {graphicx}
\usepackage {longtable}

\shortauthors{Rieke et al.}

\begin{document}

\title{Absolute Physical Calibration in the Infrared}

\author{G. H. Rieke\altaffilmark{1}, M. Blaylock\altaffilmark{1,2}, L. Decin\altaffilmark{3}, C. Engelbracht\altaffilmark{1}, P. Ogle\altaffilmark{4}, E. Avrett\altaffilmark{5}, J. Carpenter\altaffilmark{6}, R. M. Cutri\altaffilmark{7}, L. Armus\altaffilmark{4}, K. Gordon\altaffilmark{1,8}, R. O. Gray\altaffilmark{9}, J. Hinz\altaffilmark{1}, K. Su\altaffilmark{1}, Christopher N. A. Willmer\altaffilmark{1}}

\altaffiltext{1}{Steward Observatory, University of Arizona, 933 North Cherry Avenue, 
Tucson, Arizona 85721}
\altaffiltext{2}{currently at UC Davis, One Shields Ave., Davis, CA 95616} 
\altaffiltext{3}{Department of Physics and Astronomy, Institute for Astronomy, K.U.Leuven, Celestijinenlaan 200B, 
3001, Leuven, Belgium}
\altaffiltext{4}{Spitzer Science Center, California Institute of Technology, Pasadena, CA 91125}
\altaffiltext{5}{Harvard-Smithsonian Center for Astrophysics, 60 Garden Street, Cambridge, MA 02138}
\altaffiltext{6}{Department of Astronomy, California Institute of Technology, MC 105-24, Pasadena, CA 91125}
\altaffiltext{7}{Infrared Processing and Analysis Center, California Institute of Technology, Pasadena,
CA 91125}
\altaffiltext{8}{currently at Space Telescope Science Institute, 3700 San Martin Drive, Baltimore, MD 21218}
\altaffiltext{9}{Department of Physics and Astronomy, Appalachian State University, Boone, NC 28608}

\begin{abstract}

We determine an absolute calibration for the MIPS 24$\mu $m band and 
recommend adjustments to the published calibrations for 2MASS, IRAC, and IRAS  
photometry to put them on the same scale. We show that consistent results 
are obtained by basing the calibration on either an average A0V star 
spectral energy distribution (SED), or by using the absolutely calibrated 
SED of the sun in comparison with solar-type stellar photometry (the solar 
analog method). After the rejection of a small number of stars with anomalous 
SEDs (or bad measurements), upper limits of $\sim $ 1.5\% (rms) are 
placed on the intrinsic infrared SED variations in both A dwarf and 
solar-type stars. These types of stars are therefore suitable as 
general-purpose standard stars in the infrared. We provide absolutely calibrated SEDs for a 
standard zero magnitude A star and for the sun to allow extending this work 
to any other infrared photometric system. They allow the recommended calibration to be 
applied from 1 to 25$\mu $m with an accuracy of $\sim$ 2\%, and with even higher accuracy 
at specific wavelengths such as 2.2, 10.6, and 24$\mu $m, near which there 
are direct measurements. However, we confirm earlier
indications that Vega does not behave as a typical A0V star between the visible 
and the infrared, making it problematic as the defining star 
for photometric systems. The integration of measurements of
the sun with those of solar-type stars also provides an accurate estimate
of the solar SED from 1 through 30$\mu$m, which we show agrees with
theoretical models.

\end{abstract}


\section{Introduction}

For many applications of astronomical photometry, an accurate absolute 
calibration in physical units and at multiple wavelengths is critical. 
Previous calibrations have made use of two basic approaches (Rieke et al. 
1985). In direct calibrations, measurements are made of celestial sources in 
ways allowing the signals to be compared directly (if sometimes through a long 
chain of measurements) with signals from calibrated emitters. For indirect 
calibrations, direct calibrations at wavelengths well removed from those of 
interest are interpolated or extrapolated through physical modeling of 
astronomical sources. For direct calibrations, rigorous error analysis is 
possible, although there is always a risk of systematic terms that are not 
captured. Error analysis is far more difficult for indirect calibrations, 
since the systematic errors in theoretical modeling are often not apparent 
and there is usually no rigorous way to quantify the errors. Therefore, indirect 
calibrations must be used with caution until they have been 
confirmed by other indirect approaches or, better, by direct ones. 

Infrared calibrations have often included some aspect of indirect 
calibration by extrapolating from the high quality visible direct 
calibrations. There are now high quality direct calibrations in the 
infrared, so the extrapolations can be tested. In this work, we start with 
the infrared calibrations, test their internal consistency, and then examine 
the consequences for extrapolation from them into the visible. This approach 
is preferred in principle because stellar behavior is relatively simple in 
the infrared (e.g., small temperature uncertainties have little effect on 
the shape of a nearly Rayleigh-Jeans spectrum for stars of the temperatures 
considered here). 

The 24$\mu $m band of the Multiband Imaging Photometer for Spitzer (MIPS) 
achieves high photometric accuracy - with typical errors an order of 
magnitude smaller than those previously achieved at similar wavelengths 
(Rieke et al. 2004; Engelbracht et al. 2007). The Infrared Array Camera
(IRAC) also is providing a large body of accurate and homogeneous photometry
in the 3 - 8$\mu$m range (Fazio et al. 2004; Reach et al. 2005). 
In addition, a new absolute calibration of 
unprecedented accuracy is now available in the thermal infrared (Price et 
al. 2004). These advances make it both desirable and feasible to establish a 
very accurate calibration of the MIPS photometry at this wavelength and to 
provide guidelines to tie it in consistently with photometry at shorter 
wavelengths. 

Bessell (2005) has reviewed photometry in general, but with only a few 
comments on the infrared. Price (2004) has reviewed infrared calibrations. 
His review concentrates on a huge body of work by Cohen, Walker, Witteborn, 
Price, and other co-workers on this topic. The review states that 
the results are "in substantial disagreement with previous direct 
calibrations," thus leaving open the question of possible undetected 
systematic errors. In addition, the review points out a number of 
discrepancies between the measured properties of the sun (e.g., Thuillier et 
al. 2003) and measurements of solar analog stars, which indicate possible 
issues in the photometric system. 

The availability of large and homogeneous 
sets of data such as the Hipparcos photometric catalog (Perryman et al. 
1997) and the 2MASS survey (Cutri et al. 2003) opens new possibilities to 
probe these issues and to improve calibrations, as well as providing a solid 
foundation to extend calibrations uniformly over the entire sky. 
Doing so is the goal of this paper.  
The paper is organized as follows. In Section 2, we review measurements of 
the absolute flux from Vega at 2.22 and 10.6$\mu $m, and the extrapolation 
of these measurements to the MIPS effective wavelength of 23.675$\mu $m. In 
Section 3, we discuss an independent approach to calibration using the solar 
analog method introduced by Harold Johnson (1965a). In Section 4, we show 
explicitly the discrepancy in the Vega-based visible calibration and 
the infrared one. In Section 5, we 
recommend adjustments to the absolute calibration of other photometric systems to bring them 
into agreement with the work reported here. The paper 
is summarized in Section 6. Those not wishing to plow through the details 
can go to that section for a summary of the useful results, 
which include recommendations for an absolute calibration accurate to
2\% or better across the near and mid infrared.

\section{Direct Infrared Calibrations of "Vega"}

\subsection{Zero Point}

Traditionally, absolute calibration systems have been referred to a "zero 
point" of a magnitude system, usually defined by the spectral energy 
distributions of A stars. The Johnson/Arizona system defined the zero point 
as an average of the colors of a number of A stars, not just those of Vega, and as a result the 
magnitude of Vega is slightly positive ($\sim $ 0.02) at most bands. This 
situation has caused confusion because some have used Vega by itself to 
define zero magnitude, introducing a small offset in nominally similar 
systems. Further complications arise for the wavelengths of interest for 
this paper because of the contribution of the Vega debris system to the 
fluxes from this star beyond 10$\mu $m (e.g., Aumann et al. 1984). In 
addition, Vega is a rapidly rotating pole-on star with a significant 
temperature gradient ($\sim $ 1500K) from its pole to its equator (Gulliver 
et al. 1994; Aufdenberg et al. 2006), so its spectral energy distribution 
(SED) can differ from conventional models that assume a non-rotating star 
with a single surface temperature. 

Nonetheless, there is a large body of data based on Vega. Fortunately, in 
the infrared the differences in photospheric colors due to the rapid 
rotation and other modeling uncertainties are small (see Price 2004, Figure 
5 and Section 4 below). We therefore quote the results relative to the flux 
density of "Vega", a mythical star with a spectral energy distribution given 
by a Kurucz 1993 model spectrum of an A0 star (Kurucz 2005) with 
T$_{eff}$ = 9550, log g = 3.95, log z = -0.5, and normalized to measurements of 
Vega that are corrected, if necessary, for the infrared excess from 
the debris disk around this star. This convention maintains 
continuity with the large existing
body of infrared photometry. We have compared this model with the 2003 version 
(Kurucz 2005); the differences at photometric resolution and at 
wavelengths longer than 1$\mu $m are less than 0.1\%, while the V - K color 
is 0.2\% bluer with the newer model, again a negligible difference for our 
purposes. Thus, the Kurucz Vega models provide a stable reference baseline 
in the infrared. We provide the specific model we have used in electronic form
so it can be utilized explicitly in future work or in adjustments to the
calibration (Appendix A). In the following, we place "Vega" in 
quotes because the zero point of the system is defined by an 
idealized version of this star.

\subsection{Mid-Infrared}

In this subsection, we discuss absolute calibrations near 10$\mu $m and use 
them to establish a "best" value of 35.03 $\pm$ 0.30 Jy for the 
monochromatic flux density of "Vega" at 10.6$\mu $m. We also show that the 
absolute measurements at 21$\mu $m are consistent with this value and derive 
a monochromatic flux density for "Vega" at the effective or mean 
wavelength of the MIPS 24$\mu $m band.

For reasons given in the introduction, we place highest weight on direct 
calibrations in the thermal infrared. Such measurements are summarized in 
Table 1. The two most accurate sets of measurements - Rieke et al. (1985) 
and Midcourse Space Experiment (MSX) - have estimated errors of 3\% (Rieke 
et al. 1985) or $\sim$ 0.6\% (Price et al. 2004) near 10$\mu $m and are in 
excellent agreement to within these errors. Rieke et al. (1985) review 
previous work (Becklin et al. 1973; Low and Rieke 1974) and show it agrees 
closely with their calibration, well within the errors of order 7\% quoted 
for the earlier work. The solar analog calibration in this work agrees to 
within $\sim$1.5\%, as does another calibration conducted in support
of the Infrared Space Observatory (ISO) (both to be discussed below). 
The consistency of these independent determinations indicates that 
there are no major systematic errors. 
At 21$\mu $m, the agreement between the MSX calibration and previous work is 
also excellent although the errors estimated for the Rieke et al. (1985)
measurements are about 7\%. Even though Rieke et al. reported a direct 
measurement at this wavelength, they based the recommended calibration on an 
extrapolation from the 10.6$\mu $m calibration because they felt it was more 
accurate.

To obtain the best possible calibration in the mid-infrared, we therefore 
need an optimum way to combine the measurement of Rieke et al. (1985), the 
three near 10$\mu$m from Price et al. (2004), and the solar analog determination from this 
work. We do so at a wavelength of 10.6$\mu $m, using the "Vega" 
spectral energy distribution as the means to interpolate or extrapolate to 
the same wavelength and thus to relate the measurements to each other. That 
is, the calibration is relative to the normalization in Price (2004), Figure 
5. However, this figure plots the calibration of Rieke et al. (1985) 
incorrectly. The figure shows the proposed zero point of the photometric 
system as if Vega were zero magnitude but Rieke et al. (1985) set Vega to a 
magnitude of 0.02. That is, the measurement of Vega should be 0.02 lower 
than plotted, and hence in even better agreement with the MSX values than 
indicated (Price, private communication). 

Various ways to combine the data are indicated in Table 1. For extrapolating 
to other wavelengths, we define the "monochromatic" flux 
density to be proportional to the average over a 1\% spectral bandwidth of 
$\nu ^{-2}$ f$_{\nu} $ = $\lambda ^{4}$ f$_{\lambda}$. For example, we
determine the MSX value starting from the N band flux density in 
Tables 1 and 2 of Cohen et al. (1992), extrapolated to
10.6$\mu$m according to the "Vega" SED. We used the standard deviation of
the biases in bands A, C, and D in Table 9 of Price et al. (2004) to
estimate a 1.1\% rms scatter and then combined this value with the quoted
uncertainties to compute a weighted average of the biases, which was
used to adjust the Cohen et al. value for the flux density. 

In the following 
section, we use the Rieke et al. and MSX weighted average as the basis to 
project the 10.6$\mu $m calibration back to 2.22$\mu $m, for comparison with 
direct measurements there. In Section 3.3.3, we use the direct calibration 
at 2.22$\mu $m, the IRAC 8$\mu $m measurements of solar-type stars, and the 
spectral energy distribution of the sun to obtain the independent new 
calibration listed as solar analog (this work). 

We also tabulate the calibration of 
Hammersley et al. (1998), conducted in support of the ISO mission. They 
used the 1993 Kurucz model of Vega to extrapolate from K-band measurements, 
and quoted a photospheric flux density from this model at 10.47$\mu$m.
We have corrected their value to 10.6$\mu$m and reduced it by 1.29\% to
correct for the K-band excess of Vega (Absil et al. 2006). It then agrees
excellently with the other determinations. Hammersley et al. show that their measurements
of stars within this system are also in close agreement with previous
measurements of the same stars. We have not included this calibration in
our average because it is not clear how to evaluate the errors, but
they would appear to be similar to those of most of the other entries.

All of the approaches are 
consistent with the adopted value of 35.03 Jy for the monochromatic flux 
density of "Vega" at 10.6$\mu $m. We quote a slightly increased error from 
the pure weighted average value to allow for any residual systematic errors. 
A measure of the degree of agreement is that the calibration ignoring the 
MSX result is accurate to 2\% and agrees with the MSX calibration to within 
1\%. 

We can now 
compute that the corresponding "Vega" flux density at 23.675$\mu $m (the mean 
wavelength of the MIPS band) is 7.15 $\pm$ 0.11 Jy, where we have 
assigned a 1.5\% error to allow for any issue in propagating the calibration 
to 24$\mu $m. MSX also obtained a calibration at 21.3$\mu $m, which is 
equivalent to a "Vega" flux density at 23.675$\mu $m of 7.19 $\pm$ 
0.11 Jy, i.e., is fully consistent with our extrapolation from 10.6$\mu $m. 
We take the average value of 7.17 $\pm$ 0.11 Jy = 3.835 X 10$^{-18}$ 
W cm$^{-2}$ $\mu$m$^{-1}$ as the "best" 
estimate; we have not decreased the error bar in the average because the two 
determinations are not completely independent. 

\subsection{Near Infrared}

This subsection addresses tying the absolute measurement at 10.6$\mu $m to 
direct calibrations of Vega near 2$\mu $m. Since the high-weight 
calibrations are at 2.20 and 2.25$\mu $m (Blackwell et al. 1983; Selby et 
al. 1983; Booth et al. 1989), we correct all of them to 2.22$\mu $m; this 
wavelength has the further advantage that it is well-removed from strong 
spectral absorptions in both A and G stars. Figure 5 of Price (2004) 
demonstrates that the predictions of A-star models are very similar in 
spectral shape between 2 and 24$\mu $m, and we obtain a similar result 
comparing the 1993 and 2003 Kurucz models for Vega. At shorter wavelengths, 
there can be slight deviations of models. However, 
the comparison of calibrations at 2 and 10$\mu $m and interpolations within 
this range should be robust. 

The flux density from "Vega" at 2.22$\mu $m is predicted to be 649 Jy, using 
the Kurucz model normalized at 10.6$\mu $m to the "Rieke plus MSX" value. We 
assign a 1.5\% error to this estimate to include any issues in propagating 
it from 10.6$\mu $m. This value is compared with the direct measurements of 
Vega by Walker (1969), Selby et al. (1983), Blackwell et al. (1983), and 
Booth et al. (1989) in Table 2. This list of measurements represents all the 
absolute calibrations in the literature except those based in some way on 
extrapolating the Vega spectrum (or, in the case of Campins et al. (1985) 
that are revised in this work). All the reported measurements in this table 
have been corrected to a wavelength of 2.22$\mu $m according to the Kurucz 
model spectrum of Vega. 

Absil et al. (2006) report interferometric measurements of Vega at 2$\mu $m 
that show it has an excess of 1.29\% $\pm$ 0.19\% within a field of 
diameter 2", presumably due to a hot inner circumstellar disk (see also 
Ciardi et al. 2001). We have corrected the direct measurements downward by 
1.29\% to remove the effects of this disk. It is unlikely that the output of 
the disk exceeds this value significantly. For example, the excess of $\sim 
$ 4\% observed by MSX in bands C and D at 12.1 and 14.7$\mu $m (Price et al. 
2004, Table 4) and an upper limit to the color temperature for the excess of 
2000K (compare Absil et al. (2006)) predict an excess above the photosphere 
of 0.9\% at 2.2$\mu $m. This rough upper limit is very close to the measured 
excess at this wavelength, i.e., there is no missing flux that might lie outside
the 2" field. Although there are accurate direct measurements of the flux from Vega at other
infrared wavelengths (e.g., Mountain et al. 1985), the lack of detailed understanding
of the behavior of the circumstellar material makes it problematic to
interpret these measurements to the level of accuracy that can be achieved at 2$\mu$m. 

Nonetheless, because the extent and other aspects of the correction may be 
more uncertain than indicated by the nominal error bar, we have assigned an 
error of one percentage point to the correction. The resulting value is 645 
$\pm$ 15 Jy for the directly measured flux density from the Vega 
photosphere at 2.22$\mu $m. The agreement with the value extrapolated from 
10.6$\mu $m is virtually perfect. We combine the two values to determine
a "best" calibration. 

The central result from this section is captured in Tables 1 and 2. The 
theoretical spectral energy distribution of "Vega" links the accurate 
absolute measurements of this pseudo-star at 2.22 and 10.6$\mu $m well. Because 
consistent values are obtained in distinct calibrations with completely 
different chains of measurements and assumptions, this agreement appears not 
to be undermined by any plausible systematic errors. The uncertainties in 
our final derived flux densities for this "star" are less than 1.5\% at both 
wavelengths.

\section{An Independent Verification via the Solar Analog Method}

\subsection{Spectral Energy Distribution of the Sun}

A further test of the "Vega"-based calibration is to show that an 
approach based on a different stellar type is consistent with it. To 
generate this new calibration, we use the solar analog method (Johnson 
1965a). That is, we take the absolutely calibrated measurements of the sun 
and apply them as colors to other solar-type stars. This subsection 
discusses the various forms of solar SED that we examined. 

\subsubsection{Measurements of the Solar SED}

We take the solar SED for 0.2 to 2.4$\mu $m from Thuillier et al. (2003 and 
private communication). This work supplants previous work, although it is 
generally consistent with the earlier measurements as summarized by 
Labs and Neckel (1968, 1970), to within a few percent.

Vernazza et al. (1976; hereafter VAL)
provide a careful, critical assessment of 
the measurements at longer wavelengths. It is 
concluded that the data out to 12$\mu $m are of high 
quality. Specifically, Saiedy and Goody (1959) estimate that their 
measurement at 11.1$\mu $m is accurate to 0.7\%, standard error. Saiedy 
(1960) estimates the standard errors at 8.63 and 12.02$\mu $m to be 0.9 and 
1.8\% respectively. Beyond that wavelength, the measurement accuracy 
decreases substantially. VAL made adjustments of $\sim 
$4\% in some measurements to improve the apparent agreement. The quoted 
errors are also a few percent. 

VAL described these results with a semi-empirical model \footnote{Adjustments 
were made in this model by Maltby et al. (1986), and by Fontenla et al. (2006). 
The 2006 paper tabulates the latest version of Model C for the average 
quiet sun. These changes do not modify the computed infrared spectrum 
significantly.}. The model and the final adjusted set of measurements 
from VAL are described well by a functional fit due to Engelke (1992), which we will use 
to represent the long wavelength observations for the rest of this paper.
We assign a 5\% uncertainty to the measurements beyond 12$\mu$m 
as represented by the Engelke (1992) function. 

\subsubsection{Models of the Solar SED}

We have compared the measurements with two photospheric models that
predict the solar spectral energy distribution. They are 
described in more detail here.

The model of Holweger \& M\"uller (1974, hereafter 
HM74) is based on a local thermal equilibrium (LTE) analysis of solar 
line observations. The thermal structure of 
the temperature minimum region and the chromosphere 
lying above the photosphere is controversial. The solar temperature 
structure appears to have a minimum of $\sim $ 4000K near a depth of 500km 
in the photosphere, with an overlying 1500km thick, 7000K plateau in the 
mechanically heated chromosphere. However, the analysis of carbon monoxide 
(CO) lines indicates a very cool brightness temperature ($\sim $ 3700K) at 
the extreme edge of the solar disk, where the slanted line of sight probes 
into the low chromosphere (see Ayres et al. 2006). In a study of the CO 
fundamental lines, Harris et al. (1987) concluded that the HM74 photospheric 
model was consistent with the visible continuum center-limb behavior 
and the properties of the CO fundamental spectrum. Using visible continuum 
intensities and center-limb behavior in combination with the CO center-limb 
behavior, Ayres et al. (2006) recently re-determined the solar photospheric thermal 
profile, which also closely resembles the HM74-model structure. 

Theoretical IR spectra were calculated using the HM74 model and the TurboSpectrum program 
described by Plez et al. (1993), and further updated. The program treats the 
chemical equilibrium for hundreds of molecules with a consistent set of 
partition functions and dissociation energies. Solar abundances from Anders 
\& Grevesse (1989) have been assumed, except for the iron abundance, 
$\varepsilon $(Fe) = 7.51, which is in better agreement with the meteoritic 
value. The continuous opacity sources considered are H$^{-}$, H, Fe, (H+H), 
H$_{2}^{+}$, H$_{2}^{-}$, He I, He I$_{ff}$, He$^{-}$, C I, C II, C 
I$_{ff}$, C II$_{ff}$, C$^{-}$, N I, N II, N$^{-}$, O I, O II, O$^{-}$, 
CO$^{-}$, H$_{2}$O$^{-}$, Mg I, Mg II, Al I, Al II, Si I, Si II, Ca I, Ca 
II, H$_{2}$(pr), He(pr), e$^{-}_{sc}$, H$^{-}_{sc}$, H$_{2} \quad _{sc}$, 
where `pr' stands for `pressure induced' and `sc' for `scattering'. The main 
continuous absorber in the IR is H$^{-}_{ff}$, for which the absorption 
coefficients of Bell \& Berrington (1987) were used. 
For the line opacity, we used the atomic and molecular database created by 
Decin (2000) and Decin et al. (2003). The main molecules included are CO, SiO, 
H$_{2}$O, OH, NH, CH, CN, and HF. A full spectrum from 2 to 200$\mu $m was 
generated at a resolution of $5 \times 10^{-5}$ $\mu$m.  

The Fontenla et al. (2006) model assumes a temperature minimum of 4500K and 
is computed only to a temperature of 5374K in the low chromosphere. It has 
been found that the higher chromospheric layers do not affect the spectrum 
between 1 and 100$\mu $m. The densities are computed from hydrostatic 
equilibrium and charge conservation, and the calculation assumes local 
thermal equilibrium (non-LTE effects have negligible influence on the 
infrared spectrum). Further details are given by Fontenla et al. (2006).

\subsubsection{Synthetic Solar Colors} 

All of the approaches discussed above for describing the measurements 
or modeling the solar output agree excellently in the 2 to 4$\mu$m region.
There are modest divergences at longer wavelengths (particularly beyond 10$\mu$m),
but still generally within the expected errors. 

We will therefore use photometry of solar-type stars to help decide 
among the possibilities. The first step is to compute synthetic colors for comparison with the stellar
photometry. We use the Thuillier et al. (2003) solar
spectrum at wavelengths short of 2.4$\mu$m and the Engelke function beyond 
2.4$\mu$m to represent the measurements of the sun . They join in a consistent manner
with no renormalization, as shown in Figure 1. Synthetic colors are
also computed directly for the models. To compute the K-band signal for HM74, we continued the model
to wavelengths short of 2$\mu$m based on the Thuillier et al. measurements.
The exact form of this continuation has only a modest influence on the K photometric
color. 

In the following sections, we will use near infrared 
magnitudes as defined by the 
2MASS system. To determine synthetic colors for the sun, we took the 
relative response functions from the 2MASS web site (originally from 
Cohen et al. 2003). The information on
the IRAC 8$\mu$m band is from the IRAC Data Handbook. 
The MIPS 24$\mu $m relative spectral 
response is taken from the MIPS Data Handbook. We obtained
the response of the V filter from Johnson (1965b) and multiplied it by a 
function proportional to wavelength to convert it 
into a relative response function. We convolved the Kurucz A star
spectral energy distribution and the various models for the sun with 
these functions. The relative responses to the A-star and solar spectra can then be used 
to calculate the synthetic colors provided in Table 3.

\subsection{Photometry of Solar-Type Stars}

For comparison with the synthetic colors, we determine accurate averages 
for the measured stellar colors in this 
subsection. We emphasize the use of large sets of homogeneous measurements 
(Hipparcos, 2MASS, and homogeneous data sets with MIPS and IRAC). 
An essential aspect of these comparisons is the linearity of MIPS and IRAC
over the range of the observations, which we demonstrate in Appendix B is adequate
for our needs.

\subsubsection{K$_{S}$ - [8] Color of Solar-Type Stars and a Solar Analog Calibration}

We first discuss our procedures at 8$\mu$m (further details are in Appendix C). 
The existing calibration of the IRAC photometry is 
relative to "Vega" as the zero point, as described by Reach et al. (2005). We used their stated 
zero point to compare their assumed flux density for "Vega" with ours. We 
first compute the monochromatic flux density at 7.872$\mu $m, and then apply 
their recommended color correction of 1.042. Extrapolating from the 
2.22$\mu$m calibration we find a value 1.2\% brighter than theirs, while 
extrapolating from 10.6$\mu $m we find one that is 1.7\% brighter. Thus, to 
put their calibration on the same overall scale as is recommended here, an 
upward adjustment of about 1.5\% is required. 

We have used the Formation \& Evolution of Planetary Systems (FEPS) 
Delivery 3 data products (NASA/IPAC Infrared Science Archive 2007) for a 
solar analog calibration at 8$\mu $m. We also need stars 
with very homogeneous near infrared photometry. We therefore 
required that each star have 2MASS measurements of "A" quality in all three 
bands (JHK), and that they all be measured in the "Read 1" mode (see Appendix D). 
In addition, we used the Hipparcos photometry at V to compute V - K colors and rejected any 
star departing by more than 0.10 magnitudes from the standard color. 
The final sample is listed in Table C1, and the IRAC reductions
are described in Appendix C. 

To look for intrinsic scatter in the stellar colors, we averaged the
2MASS J, H, and $K_S$ measurements to a single "$SuperK_S$" value (see Appendix C). 
We have computed the ratio of 2.2$\mu $m to 8$\mu $m flux densities for the 
stars in Table C1, using the $SuperK_{S}$ magnitudes. We find that the rms 
scatter is only 2.05\%. This value is smaller than would have been predicted 
from the combination of the uncertainties in the $SuperK_{S}$ magnitudes and 
in the IRAC 8$\mu $m flux densities. We conclude that the photometry is well 
behaved and that all of the stars have very similar SEDs between 2 and 8$\mu 
$m. (An exception arises at the CO fundamental and first overtone bands 
(bandheads at 4.6$\mu $m and 2.3$\mu$m respectively) 
due to variations in the absorption strength; these regions are not 
probed by the photometry we have used for calibration.) 

Our empirical solar SED model (Thuillier plus Engelke) lies a 
factor of 1.026, or 0.028 magnitudes, above the 
"Vega" SED, at 7.872$\mu $m, if they are set equal at 2.22$\mu $m. The 
average value of K$_S$ - [8] from the measurements of the solar-type stars in Table C1 is a 
factor of 0.986 below the Thuillier/Engelke solar SED. The errors in the solar measurement should be
small in this region, of order 1\% (VAL). We can derive an 
independent calibration by normalizing to the results at 2.22$\mu $m based 
on the direct measurements of Vega in Table 2 (i.e., excluding the 
extrapolation from 10.6$\mu $m). The resultant
calibration is entered in Table 1; we have 
assigned a 3\% error, based on the error in the 2.22$\mu $m calibration and 
the uncertainties in propagating it to 10.6$\mu $m. It agrees well with
the other calibrations.  

We conclude that a completely independent check of the linkage of the 2.22 
and 10.6$\mu $m calibrations via solar type stars agrees excellently with 
the results from direct absolute calibrations and the "Vega" SED at both 
wavelengths.

\subsubsection{Zero Color at 24$\mu$m}

We now extend the solar analog method to 24$\mu$m, to test the various
alternatives for the solar SED in this spectral region. 
We determine zero color at 24$\mu$m by averaging the measurements of a large number of A 
stars, similar in spirit to the original A-star-based zero point (Johnson and Morgan 1953). In 
our situation, the approach has a number of virtues. First, because it uses 
averages of many measurements, it achieves high accuracy in the comparison. 
Second, peculiar behavior by a few stars will have little influence on the 
results, and sufficiently peculiar stars stand out and can be rejected as 
outliers to make their influence disappear entirely. Third, the procedure 
removes our dependence on previous calibrations. 

We selected the sample of stars to use at 24$\mu$m from Su et al. (2006) supplemented by stars 
in the MIPS calibration program. We eliminated all stars with indications of 
excess emission at either 24 or 70$\mu $m (Su et al. (2006) show that $\sim$ 32\% of typical A stars have excess emission at 24$\mu $m). To guard against 
subtle excesses, we also eliminated stars younger than 200 Myr, since the 
excesses at 24$\mu $m decay roughly as time/150Myr (Rieke et al. 2005). 
The stars are listed along with their key parameters in Table C2 (Appendix C). 
Appendix C also describes our
reduction procedures at 24$\mu$m in detail.

We have fitted a Gaussian to the distribution of $SuperK_{S}$ over 24$\mu$m flux 
density ratios (normalized to one) for our A star sample. 
The standard deviation is 0.048 (we have excluded HD 172728 
from the fits because its low values for two measurements imply a possible 
problem with the 2MASS measurement and also the two stars with the highest 
ratios, HD 11413 and HD 92845, since they may have weak excess emission). By 
taking the quadratic difference of the fitted standard deviation and the 
estimated errors in the 24$\mu $m and $SuperK_{S}$ values, we find a residual 
uncertainty term of 3\%. This value is an upper limit to the intrinsic 
star-to-star rms differences in K$_{S}$ - [24] photospheric color. 

Because the intrinsic scatter appears to be small, we assume that the scatter in the colors
is dominated by measurement errors and it is appropriate to reduce 
the uncertainties by averaging. We 
found that attempting to correct the K$_{S}$ measurements for extinction had 
a negligible effect on the average (0.004 magnitudes) and increased the 
scatter, so we have used the uncorrected K$_{S}$ values. In an arbitrary 
normalization that brings the value of the flux density ratios close to 1 
(and will be preserved for a similar calculation for solar type stars) the 
average ratio of K$_S$ to 24$\mu$m flux densities is 0.964 $\pm$ 0.008.

\subsubsection{24$\mu$m Measurements of Solar-Type Stars}

We now apply the identical procedures to 24$\mu$m measurements of 
a suite of solar-type stars. Our sample is drawn largely from the FEPS
program, Delivery 3 data products. 
It is listed in Table C3 and Appendix C gives the details of our reductions. 
To guard against excess emission, we have only included stars older 
than one Gyr as determined by Wright et al. (2004).

We computed $SuperK_{S}$ magnitudes for these stars (see Appendix C). A Gaussian fitted to 
the resulting distribution of $SuperK_S$ to 24$\mu$m flux ratios 
(normalized to 1) has a standard deviation of 0.034. 
A quadratic subtraction of the estimated measurement errors from the fitted 
standard deviation leaves less than 1\% for the intrinsic scatter due to 
variations in the stellar SEDs. 

For the calibration calculation, we reject the two lowest and two highest 
values. The average normalized ratio of $K_S$ to 24$\mu $m flux density is 
1.005 $\pm$ 0.007. 
The ratio of the two averages for A and solar-type stars, 1.042, is then 
the color in flux units of a solar-type star relative to the A-star zero 
point. It is equivalent to a color in magnitudes of 0.045 $\pm$ 
0.011 in the sense that the solar-type stars are redder than A0V stars. 
As shown in
Table 3 and Figure 2, the resultant value for the sun at 24$\mu$m is 5\% above the
Engelke function, and we have assigned an error of 5\% to this function
at these wavelengths. Hence, the agreement is within the errors. However,
the color of the solar-type stars is well enough determined to suggest
that the Engelke function is 3 to 7\% too blue relative to the
true solar SED.

\subsubsection{Solar Analog Calibration at V, J, H, and K}

We can test the V, J, H, and K-band calibration by checking to see if we get the 
correct colors for the sun. From the synthetic colors, we find V - K$_{S}$ = 
1.568. The error is a combination of that in the K absolute calibration 
and in the measurements of the sun. 
>From Table 2, we take the first error category to be 1.2\%. The second class of errors is 
quoted by Thuillier et al. (2003) as 1.1, 0.8, 0.65, and 0.6\% 1-$\sigma$ 
respectively at 0.95, 1.5, 1.1, and 2.5$\mu $m. We therefore quote a net 
error of 2\%. Similar errors should hold for the other bands.

In principle, this solar color should agree with the colors of similar 
stars. The V and K$_{S}$ colors of Vega and solar analog stars are tied 
together by accurate direct calibrations. However, the J and H 2MASS 
measurements are determined by color transformation and interpolation of the 
direct calibrations. For 2MASS observations of these relatively bright 
stars, it is possible that there are residual errors at 
the 1 to 2\% level. Therefore, rather than 
assuming the 2MASS color zero points, we determined the zero points for the 
J and H bands by averaging measurements of a large number of A0V stars also 
measured in the Read 1 mode. Our procedure is discussed in Appendix D. 
We then corrected the catalog solar analog 
colors to these zero points. 

Our sample of solar-type stars is largely from the NStars
compilation (Gray 2007). We fitted the trend of colors with temperature 
and used the fit to adjust them all to match the color expected for a 
star with a temperature of 5778K (the effective temperature of the sun) -
see Appendix D. 
The final average V - K$_{S}$ color of 1.545 $\pm$ 0.015 is compared with 
the V - K$_{S}$ of the sun in Table 3. 

Our value of $<$V - K$_{S}$$>$ = 1.545 for the average of solar-type stars 
differs substantially from standard tabulations such as 1.46 in Tokunaga 
(2000), as well as other determinations such as that of Holmberg et al. 
(2006). Part of these discrepancies may be traceable to the 0.045 magnitude 
correction implied for the SED anomalies of Vega, but another important 
contributor is possible discrepancies in translating the solar temperature 
into the stellar temperature scale (Holmberg et al. 2006). There is also significant 
scatter in assigned temperatures within a spectral type in the Nstars 
compilation (Gray 2007), equivalent to at least one spectral subtype. Roughly speaking, 
a shift of 10K in the temperature of a solar-type star shifts its V - K by 
0.01. 

For another comparison, we used a very carefully compiled and clearly 
described set of colors provided by Bessell et al. (1998): 
see Appendices A - C of their paper. They define a magnitude system in 
which V - K = 0 for Vega, and have forced the calibration to fit both 
this color definition and a similar model for "Vega" as the one used in 
this paper. By forcing the "Vega" model 
to fit the infrared calibrations, we find a system in which V - K = 0.045 
for Vega the real star (see below). From Carpenter (2001), we find an additional 
adjustment of 0.018 from the Glass/SAAO system to 2MASS, or a total of 
0.063. Table 4 allows comparison of the various estimates of V - K$_{S}$ 
with the Bessell et al. (1998) results corrected to the same basis as ours. 
Details regarding the first four entries in this table can be found in 
Bessell et al. (1998). We have also entered the measurements from Table 3 to 
demonstrate the good agreement. We show additional synthetic
colors from models computed by Casagrande et al. (2006). They
assumed a $V - K_S$ color of 0.047 for Vega, bringing their scale close to the
one we have used (with $V-K_S = 0.045$). 
The prime solar-type NICMOS calibrator, P330E, can be used as an independent 
test of these colors. We show in Table 4 the colors of P330E relative to a 
synthetic V magnitude (Bohlin et al. 2001). We have interpolated to provide 
a K$_{S}$ magnitude. All of these determinations of the solar $V - K_S$ color 
agree very well. 

We also computed J - K$_S$ and H - K$_S$ colors for the 32 solar-type 
stars listed in Table D2 with A grade Read 1 measurements in all three 
bands (this criterion eliminates HD 41330, 90508, 168009, and 186427 from 
the sample). Since these colors are relative to the 2MASS system, we 
corrected them for the slight deviations from zero color we found for A0 
stars (see Appendix D). The average colors are listed in Table 3 and plotted in Figure 2
for comparison with the solar SED. The errors are estimates from 
those quoted in the 2MASS catalog, with some allowance for systematic effects. The 
agreement with the solar colors is excellent. 

The infrared colors also 
agree well with those in Tokunaga (2000): H - K = 0.061 for us, vs. 0.05, 
and J - H = 0.326 vs. 0.32. We can also compare with Bessell et al. (1998), 
but we first transform the J - K$_{S}$ to the Johnson/Glass system by 
subtracting 0.007 magnitudes, determined from the Thuillier solar SED. 
We then find J - K = 0.38 and H - K 
= 0.054, compared with J - K = 0.38 and H - K = 0.045 from their tabulation 
of the solar analog determinations of Cayrel de Strobel (1996). 

If we predict errors from the quoted uncertainties in the 2MASS measurements, we obtain a 
predicted error for a typical J - K$_{S}$ measurement of 3.1\%, whereas the 
scatter in this color indicates a typical error of 2.3\%. This behavior is 
consistent with some degree of correlation in the 2MASS measurements, which 
is reasonable. There is no indication of scatter in the intrinsic colors of 
the stars. A similar argument indicates no detected intrinsic scatter in H - 
K$_S$.

We have made a more demanding test for the uniformity of the JHK colors of
solar-type stars. We used the accurate near infrared photometry of Kidger \&
Mart\'in-Luis (2003), which for well-measured stars has errors
of less than 1\%. We did not use other available high accuracy 
photometry compilations that concentrate on faint sources to calibrate 
infrared arrays, since these stars are more distant and subject to reddening.  
We selected the 16 stars from Kidger \& Mart\'in-Luis with listed spectral types of G0
through G5 IV or V, with $1.4 < V-K < 1.8$, with six or more measurements,
and with errors in all three bands indicated as $<$ 1\%. We then fitted a
straight line to the trends of $J-H$ and $H-K$ vs. $V-K$, finding scatter
around the line of 1.5\% in the first case and 0.9\% in the second. We used
$V-K$ instead of spectral type because we did not want the results to
be subject to type errors. In addition, Kidger \& Mart\'in-Luis (2003) include a 
number of color-color plots that show small scatter
that is independent of spectral type from A through K stars. We conclude that
G stars have intrinsic scatter in the near infrared colors of no more than about 1\%.

The "anomalously red" color of the solar spectrum as measured by Thuillier
et al. (2003) relative to such determinations as Holmberg et al. (2006) has
not been satisfactorily explained previously (see discussion
in Casagrande et al. 2006). It is comforting that, with care in
analyzing the photometric database, we have found that this color is
consistent with those of other solar-type stars, and that the scatter in
color among such stars is small. As shown in Table 3, this color is consistent
not only with the empirically measured solar colors but with the predictions
of a large number of models of the solar SED. 

\subsubsection{Spectra}

We have used the Spitzer Infrared Spectrograph (IRS) to confirm the slope 
and overall spectral behavior of the solar-type stellar SED between 8 and 
30$\mu $m. The result, reduced as described in Appendix C, is plotted in Figure 2, 
normalized to the photometric point at 8$\mu $m. If it is normalized
to the HM74 model at 8$\mu$m, the rms noise around the model continuum 
is $\sim $ 0.7\%. It therefore confirms to high accuracy the overall shape of the
SED of solar type stars as described by this model.

\subsubsection{An Empirical Solar SED}

As shown in Figure 2, the HM74 model agrees well with the measurements of both the sun and
of solar-type stars: 1.) it tracks the Engelke function closely out to about
10$\mu$m; 2.) it is consistent with the solar-type stellar calibration
at 8$\mu$m; 3.) it is also consistent with the solar-type stellar color at
24$\mu$m; and 4.) it agrees with the overall spectral shape of solar-type
stars measured with IRS between 8 and 30$\mu$m. The model also
includes a full treatment of the solar absorption line spectrum. 

We therefore adopt it to describe the mid-infrared SED of the sun. 
To normalize it to the Thuillier et al.
spectrum, we took advantage of the fact that the absolute level in the Engelke 
function agrees very closely with the Thuillier spectrum at 2.3 - 2.4$\mu $m, so 
we can use this fit as a smoothing function to help join the two spectra. 
Figure 1 illustrates how the HM74 model was joined to the Thuillier measurements. 

The remaining issue is that the CO fundamental bands are difficult to fit
{\it a priori} in models. We empirically adjusted the depth of these features
by setting the CO absorption features to be consistent with the 
spectroscopy of Wallace \& Livingston (2003). At their high spectral
resolution, there are a number of atmospheric mini-windows that allow
accurate measurement of CO equivalent widths. On average, we found that the
HM74 model had slightly weaker CO absorption than found by Wallace \&
Livingston. We therefore used the Engelke approximation between 4 and
6.5$\mu$m as a featureless continuum, normalized to the HM74 model. 
We multiplied the Engelke SED by 0.125, subtracted it from the HM74
model, and renormalized the result to the original continuum level to
bring the CO equivalent widths into agreement with those of Wallace \& 
Livingston (2003). This modified spectrum was used to replace the 
HM74 values between 4 and 6.5$\mu$m. 

Our final adopted solar spectrum is shown in Figure 2 and given numerically in Appendix A. 
The Engelke (1992) fit to the VAL model/reconciled measurements 
falls slightly below the empirical model at wavelengths longer
than about 6$\mu$m. In the IRAC 8$\mu$m band, the discrepancy is 1.4\%, 
at the outer limits of the errors. The model is therefore slightly
discrepant with our calibration, based on accurate solar measurements. 
At 24$\mu$m, the model and the photometry
differ from the solar measurements by 5\%, but here the model result is
to be preferred because of the larger errors in the solar measurements. 
In general, the model should represent the true solar SED to within
$\sim$ 2\%, an error estimated from a combination of the discrepancies
with the solar measurements and the photometric errors in the
solar-type stellar colors. 

\section{Vega at Wavelengths Short of 2$\mu $m}

To extend these procedures to the visible range, we use the 
average of our "best" calibrations at 10.6$\mu$m from Table 1
and at 2.22$\mu $m from Table 2 and the "Vega" model to predict a 
value of 3714 Jy for Vega at 0.5556$\mu $m. M\'egessier (1995) has 
summarized and reconciled various direct measurements of Vega, 
corrected to a wavelength of 0.5556$\mu $m. The preferred value for 
the reconciled measurements is 3563 
Jy, 4.2\% or 0.045 magnitudes less than we find at this same wavelength via 
the model. The net errors are only about 1\% for the infrared and 0.7\% at 
0.5556$\mu $m. This value agrees with the results of Bohlin and
Gilliland (2004), who find that the spectrum of Vega normalized at
0.5556$\mu$m and extrapolated using the 1993 Kurucz model to
2$\mu$m is 2\% fainter than the calibration of Cohen et al. (2003),
which we find in turn is 2\% lower than our calibration. 
Thus, although the theoretical "Vega" spectrum gives good 
agreement with measurements in the infrared, there is a significant 
discrepancy between the infrared and visible. This result is not new. 
M\'egessier (1995) discusses it at length, summarizing many results 
that point to the same issue. As a result, this work considers the infrared 
calibration separately from the visible one for similar reasons as discussed 
here. 

It seems likely that the departure from the model arises because Vega is a 
pole-on rapid rotator. With a measurement of the surface temperature 
distribution on the star (Aufdenberg et al. 2006), we can now address where 
its SED might depart sufficiently from the single-temperature 
models to be of concern for its use as a calibrator. The equatorial surface 
temperature is estimated to be 7900K, corresponding to type A7, which has V 
- K = 0.5, J - K = 0.09, and H - K = 0.03 (Tokunaga 2000). If we imagine 
the effective visible surface of the star to be a combination of A0 and A7 
spectral type to give a net V - K = 0.045, by interpolation we expect an 
effect of $\sim $ 0.01 magnitudes in J - K and 0.003 magnitudes in H - K to 
allow for the cooler portion of the surface (these values have little dependence
on the exact spectral types used to fit Vega). We conclude that this effect 
can be ignored in using Vega as a relative infrared calibrator 
at H band and longer wavelengths, but that measurable effects are expected 
at wavelengths short of J. 

In Appendix D, we determine a 2MASS K magnitude for Vega of -0.036 $\pm$ 0.010 
by transforming the measurements of Johnson et al. (1966) into the
2MASS system. Typical adopted
values for the V magnitude of Vega are 0.03 (Johnson et al. 1966; Gray 1998)
or 0.026 (Bohlin \& Gilliland 2004). 
The observed V-K of this star is therefore 0.062 to 0.066 $\pm$ 0.012.
By comparing the absolute calibrations at V and in the infrared, we found
V - K = 0.045 $\pm$ 0.013. An additional 0.014 $\pm$ 0.002 magnitudes 
should be added to account for the contribution of the ring, for
a net V - K = 0.059 $\pm$ 0.013. The difference in these estimates is 0.003 to 0.007 $\pm$ 0.018,
that is, it is not significantly different from zero.  
This desirable outcome would appear in part to result from the
indirect procedures used to set the zero point for most photometry
since that of Johnson. By setting the V - K colors of a large suite
of A0V stars to zero, the systems have been forced to remove any
residual anomalies due to the behavior of Vega. The result confirms our
derivation of A-star and solar-type colors in this paper under
the assumption that there are no unexpected offsets between the
visible and near infrared, despite the unexpected behavior of
Vega as the star defining the zero points.  

\section{A Consistent Calibration}

We have demonstrated that an absolute calibration can be derived between 1 and 
25$\mu $m that is consistent with all the direct 
calibration measurements, and both with A-star standards and with the solar 
spectrum as reflected by solar-type stars. However, practical photometry is 
conducted through filters of some band width, which must be taken into 
account in applying this calibration. To apply any calibration conveniently requires further 
simplification of its description through definition of a wavelength 
associated with a measurement and of an equivalent monochromatic flux 
density at that wavelength, derived from the calibration. There are a number 
of possible wavelength definitions. The simplest is the mean wavelength 
(which we also term the effective wavelength (H. L. Johnson, private 
communication)). The "nominal" and "isophotal" wavelengths are alternative ways to describe
a photometric band. Refer to Appendix E for further discussion of these issues.

The correction factors to put various sources of infrared 
photometry on the same calibration as derived in this paper are listed in 
Table 5. The existing calibrations are to be multiplied by these factors; 
for example, the IRAC calibration is slightly faint relative to the MIPS one 
and flux densities under it need to be increased by 1.5\%. Since this 
discrepancy can be traced to the flux density estimate for Vega, it should 
hold for all the IRAC bands. At 10$\mu $m, the new calibration is 0.8\% 
lower than the calibration of Rieke et al. (1985) (see Table 1). Since the 
IRAS 12$\mu $m calibration is derived directly from that of Rieke et al. 
(1985), a similar difference should hold for it (see Beichman et al. 1988). 
Cohen et al. (1992) re-calibrated IRAS at 12$\mu $m, finding a value 2.4\% 
below the Beichman et al. (1988) calibration, and thus 1.6\% lower than the 
preferred value based on Table 1. Our calibration is 2\% lower than the IRAS 
one at 25$\mu $m (see Beichman et al. 1988). Cohen et al. (1992) also 
re-calibrated this band, finding a value 6\% below that of Beichman et al. 
(1988) and 4\% below our preferred value.

With careful specification of the defining wavelengths (see Appendix E), we can now
compare the various calibrations in the near infrared. 
The 2MASS calibration at K$_{S}$ by Cohen et al. (2003) is 
2\% lower than ours (i.e., the fluxes in the 2MASS system must be increased 
by 2\% for consistency with our calibration). Appropriate calibration 
parameters for 2MASS are listed in Table 6. In addition to the relevant
wavelengths and flux densities, the table includes a color correction to
a 9550K black body to give an idea of the size of such terms for hot stars. The
tabulated number is the factor by which the defining spectral energy distribution
(flat for mean wavelength or rising in proportion to wavelength for nominal one) 
must be increased relative to the stellar zero point to give the same signal. For 
objects cool enough that the photometric bands are on the Wien side of their spectral
energy distributions, the corrections are substantially larger. 

The calibration proposed by 
Tokunaga and Vacca (2005) for the Mauna Kea Observatories near infrared 
filter set is 2.6\% lower than ours at 2.22$\mu $m. The measurement of Vega 
by Campins et al. (1985) is 1.4\% higher. The homogenized photometry 
proposed by Bessell et al. (1998) is 3.7\% = 0.039 magnitudes lower than our 
proposed calibration at 2.22$\mu $m. This shift is very likely associated 
with the red color of V - K$_{S}$ = 0.045 for Vega. Bohlin and Gilliland 
(2004) suggest using the Kurucz Vega model (T = 9550, log g = 3.95) to 
extrapolate from the V calibration into the infrared; we have shown that the 
resulting calibration will be 0.045 magnitudes lower than the direct 
infrared measurements. 


\section{Summary}

We have reviewed the calibration of infrared (1 to 25$\mu $m) photometry. 
Our most important conclusion is that there is very consistent behavior of 
solar-type and A-type stars, and that they in turn are closely consistent 
with virtually all direct calibration measurements and with models of their 
spectra. Concerns of significant inconsistencies (Price 2004; Bohlin 2007) can therefore 
be put aside, and we can proceed to develop a procedure for calibration of 
infrared measurements with assurance that there are unlikely to be serious 
undetected systematic errors. 

We have therefore established a consistent calibration across the near and 
mid-infrared spectral region (1 to 25$\mu $m). The foundation of the 
calibration is the accurate direct measurements near 2.2$\mu $m and 
particularly near 10$\mu $m. The accuracy of the absolute calibration is 
2\% or better across this entire wavelength range. We provide 
guidelines for applying it to 2MASS, IRAC, and MIPS photometry. Because of 
the overall agreement among the previous calibrations, the adjustments to 
apply to them for a fully consistent infrared calibration are small, 
generally within the stated errors.

After the rejection of a few stars with anomalous SEDs, 
upper limits of $\sim $ 1.5\% (rms) are placed on the intrinsic 
infrared SED variations in both A dwarf and solar-type stars. These types of 
star are therefore suitable as general-purpose standard stars, allowing the 
calibration to be extended readily to other photometric bands and systems. 
We provide spectral energy distributions of a fiducial A star and of the sun 
for use in extending the calibration to other systems, or for generating
fainter or brighter mid-infrared standards by extrapolation from accurate
near-infrared measurements. 

The suggested calibration is summarized in a number of tables. 
Table 1 gives the zero points at 10.6 and 23.675$\mu$m. Tables C2 and C3
provide a list of accurate measurements of A and solar-type stars at the 
latter wavelength. Table C1 provides measurements of solar-type stars at 7.872
$\mu$m. These measurements can be transferred to any wavelength
using the spectral energy distributions in Table A1. The listed stars are
bright enough that they can be measured at high signal to noise from
the ground at least through the 10$\mu$m atmospheric window, so they
provide a direct transfer between the Spitzer calibration and groundbased
observations. Zero points for 2MASS photometry
are provided in Table 6. 

In many cases, however, it is more convenient simply 
to adjust measurements using alternative calibrations to the 
consistent scale suggested in this paper. The relevant correction factors
for 2MASS K$_S$, IRAC Band 4 and IRAS Bands 1 and 2 are listed in Table 5. Corrections
to the other 2MASS and IRAC bands should be similar to those listed.  

Previous work has suggested an inconsistency between infrared and visible 
measurements of Vega, when fitted to a standard A0V star model. The improved 
accuracy of the infrared calibration, and its confirmation through many 
approaches, makes it clear that this inconsistency is real. The A0V star 
model normalized to the infrared calibration and extended to the visible 
(0.5556$\mu $m) predicts a flux density 4.2\% = 0.045 mag brighter than the 
absolute measurements of Vega near this wavelength. We have independently verified this
result by transforming the Vega measurements of Johnson et al. (1966) into
the 2MASS system, showing that Vega would be $\sim$ - 0.036 at 2MASS K$_S$. 
It is likely that the 
discrepancy has roots in Vega being a rapidly rotating star seen pole-on, so 
that its output spectrum is 
affected by the surface temperature gradient associated with the rapid 
rotation (Gulliver et al. 1994). We conclude that the V - K$_S$ color of 
Vega is about +0.045 magnitudes (plus 0.014 magnitudes to account for
the excess due to its ring at 2$\mu$m) compared with the average colors of A0V stars, 
and that this star is not suitable to define a photometric zero point because of 
its eccentric SED for its spectral type.

An important feature of our approach is its use of large, homogeneous 
databases - for example, Hipparcos and 2MASS photometry, the Nstars 
classification of nearby solar-type stars, and the FEPS and MIPS GTO samples 
of solar-type and A-type stars observed to identify debris disks. The 
homogeneity and generally high accuracy of these data support a new approach 
to calibration. Initially large samples of stars are cleaned of members with 
anomalous colors (due, e.g., to reddening or photometric errors) and the 
results of still moderately large samples are then averaged to drive the 
measurement errors to small values. In addition to providing an accurate 
calibration that does not depend on a small number of "ideal" objects such 
as Vega was once thought to be, this approach facilitates extending the 
calibration over the entire sky.


\section{Acknowledgements}

We thank Tom Ayres, Martin Cohen, Chris Corbally, Mark Kidger, Gerry 
Neugebauer, Stephen Price, Murray Silverstone, and Gerard Thuillier for 
helpful discussions. This publication makes use of data products from the Two Micron All 
Sky Survey, which is a joint project of the University of Massachusetts and 
the Infrared Processing and Analysis Center/California Institute of 
Technology, funded by the National Aeronautics and Space Administration and 
the National Science Foundation. This research also has made use of the 
SIMBAD database, operated at CDS, Strasbourg, France. It also
made use of the NASA/IPAC Infrared Science Archive. This work was 
supported through contracts 1255094 issued by JPL through CalTech
and NAG5-12318 from NASA/Goddard to the University of Arizona.

\newpage

\section{Appendix A} 

This appendix provides the reference spectral energy distributions for
the sun and Vega, from 0.2 to 30$\mu$m. A sample of the first few entries
is given in Table A1.

\eject

\section{Appendix B - Linearity}

The linearity of the IRAC measurements is discussed by 
Reach et al. (2005) and is adequate for the calibration we have derived. 
Here, we use the understanding gained regarding stellar colors to test 
whether there are any non-linearities at a level that would affect the MIPS 
measurements. Because we have concentrated on stars measured at very high signal to noise 
at 24$\mu $m and also measured in the Read 1 mode with 2MASS, the dynamic 
range of the measurements used in the calibrations in Sections 2 and 3 is 
small, only about a factor of five, and hence the demands on instrument 
linearity are modest. 

To test the linearity over a larger dynamic range, 
we use A-stars from Su et al. (2006) that pass all the tests for our 
calibration sample, except that they are too bright to be measured by 2MASS 
in Read 1 mode. Modern array detectors generally saturate on such bright 
stars (not just for 2MASS), so we depend on aperture photometry for the K band 
data. Using these measurements requires that the photometric system be 
understood well enough to transfer accurately to the 2MASS Read 1 system. We 
have taken measurements of HD 18978, HD 108767, HD 130841, HD 135742, and HD 
209952 from Carter (1990), of HD 80007 and HD 130841 from Bouchet et al. 
(1991), and of HD 11636, HD 16970, HD 76644, HD 87901, HD 103287, and HD 
108767 from Johnson et al. (1966). Transformations for the first two 
references were obtained from Carpenter (2001). For the third reference, we 
first determined a transformation into the CIT system of K$_{AZ}$ - 
K$_{CIT}$ = 0.024 from the table of bright star measurements in Elias et al. 
(1982), and then independently confirmed the transformation in Carpenter 
(2001) from the table of fainter standards. Thus, the net transformation to 
2MASS Read 1 observations is K$_{AZ}$ - K$_{2MASSR1}$ = 0.048, very close to 
the value of 0.044 derived independently by Bessell et al. (1998). The 
agreement of our Read-1 transformation with that of Carpenter (2001) 
validates our using his values for the other systems. 

Figure B1 shows the normalized ratio of 24$\mu $m to K flux density, vs. 
24$\mu $m flux density. We exclude the two lowest measurements (of HD 
172728). We also exclude the measurements of HD 47332 and HD 57336, the two 
highest measurements, because they may have faint excess emission. Since we 
have excluded the two highest and two lowest measurements, the bias on the 
results should be minimal. 

We have considered three types of nonlinearity. The first is the typical 
gradual reduction of response in a simple integrating amplifier as the wells 
fill. The second is an over-correction for a nonlinearity of the first kind, 
so it just reverses the sign of the curve. The third is having a small 
latent image under the image of the star being measured. Since we have used 
custom flat fields to remove latents, the sign of the latent image could be 
either positive (adding to the star signal) or negative (subtracting from 
it). Examination of our data indicates that we can place an upper limit of 
about 50$\mu $Jy on any residual latent image. 

As shown in Figure B1, the primary evidence for nonlinearity is a tendency 
for the bright stars to yield slightly larger signals than would be the case 
with a perfectly linear system. This offset is within the errors, 
but, if confirmed, it suggests that the correction for nonlinearity in large signals may be 
slightly too big (see Engelbracht et al. (2007) for further discussion). There is also
evidence for a slight tendency to obtain larger net slopes with increasing
integration time, an effect that would be consistent with the possible nonlinearity
(Engelbracht et al. 2007). However, at 
the level of our measurements, 10 to 50mJy, these effects are negligible. 

\eject

\section{Appendix C. Measurements with {\it Spitzer}}

This appendix discusses
our measurement procedures for the {\it Spitzer} data.
It also collects the samples of stars used for the various 
calculations in the paper, along with their key parameters.

\subsection{Procedures at 8$\mu$m}

One of us (J. Carpenter) re-reduced the measurements of our solar-type stars
to put them on the identical basis as the calibration reductions of
Reach et al. (2005). The 8$\mu$m solar analog calibration is based on the stars 
listed in Table C1. Fortunately, although some of the IRAC bands can 
have small offsets due to positioning of the measuring aperture, this effect 
is immeasurably small at 8$\mu $m (M. Silverstone, private communication), 
and we have ignored it. We thus only had to divide by the bandpass correction to 
convert to equivalent monochromatic flux density at 7.872$\mu $m. Since the SEDs of 
solar-type stars match very closely in this spectral region those of 
A stars, we applied the correction quoted by 
Reach et al. for an A1 star, 1.042. 

\subsection{Procedures at 24$\mu$m}

To achieve the most accurate possible data reduction at 24$\mu $m, we have 
used a series of custom processing steps that have been developed by the 
MIPS team (discussed in more detail by Engelbracht et al. 2007). These steps 
are applied after standard pipeline processing. First, we remove the 
artifacts due to dust particles on the instrument pickoff mirror. Special 
flat fields are constructed from all of the photometric data to identify the 
effects of these particles. Second, we compensate for the latent images on 
the 24$\mu $m array as a result of exposure to bright sources. These dark latents 
appear as regions of reduced sensitivity centered on the array positions of 
bright sources in subsequent exposures. An image of the latents with all 
sources removed is produced by median combining the entire data set with 
appropriate bright source masks. The dark latents are removed from each 
individual frame by dividing the frames by the normalized dark latent image 
prior to mosaicking.\footnote{A mean background frame determined from the masked 
data is subtracted as well. The mean of 
the subtracted values is added back to each frame. 
This step removes a small gradient likely due to scattered 
zodiacal light that depends on the position of the scan mirror. } 
Finally, mosaicked images 
are produced at the nominal 
pixel scale, 2.45". This processing is described in more detail by 
Engelbracht et al. (2007).

We used simple aperture photometry on the 24$\mu $m images. 
The tabulated data assumed 1 DN/s = 1.05 mJy/square arcsec.
The photometry was done within an aperture of radius 35", and relative
to sky measured in an annulus between radii of 40 and 50". The aperture
correction for this measurement approach was taken to be 1.084. 
We used the standard world coordinate 
system pointing information as a first guess for centering, with fine tuning with the IDL 
program "mpfit2dpeak.pro" written by Craig Markwardt. Comparing photometry 
for separate measurements of the same star (typically also in different 
observation campaigns) indicates that the end-to-end scatter (including 
instrumental instability as well as photometry errors) is less than 0.5\% 
rms (and not dependent on the brightness of the target for the range 
considered here -- further discussion in Engelbracht et al. 2007). Achieving 
this level of repeatability requires not only the special post-pipeline 
processing, but also careful standardization of photometric procedures. 
Therefore, all the results reported in this paper used the identical 
photometric procedures applied by the same person (M. Blaylock). The results
are listed in Tables C2 and C3; the maximum 
plausible errors in this band are no more than 1.5\% rms.

Our procedures have been selected to be very conservative (e.g., large 
measurement aperture) and adapted to high repeatability. Other photometry 
approaches can be tested by using archival data for our program stars listed 
in Tables C2 and C3. 

\subsection{Color Combinations}

Tables C1 - C3 list the 2MASS K$_{S}$ value and a second K$_S$ magnitude. 
$SuperK_{S}$ is a combination 
of the 2MASS J, H, and K$_{S}$ measurements used to measure the scatter in 
K$_{S}$ - [24] color among the stars. We used the standard colors for the spectral type 
of the star (Tokunaga 2000) to convert the J and H measurements to 
equivalent K$_{S}$ ones. The typical distance to one of the A stars is about 
100pc, so reddening may be significant. We estimated the reddening from the 
standard V-K color for the spectral type of the star (Tokunaga 2000) and the 
extinction curve of Rieke and Lebofsky (1985), and corrected all the J, H, and 
K measurements accordingly (the 24$\mu $m extinction is less than 1\% even 
for the most obscured of the stars). The $SuperK_{S}$ magnitude 
is the weighted average of all three extinction- and color-corrected measurements. 
Because the colors of A stars are close to zero already, minor differences 
in photometric system have little influence on this conversion. The 
small offsets in the 2MASS J - K and H - K colors discussed in Appendix D have 
a negligible effect on our calculations here, because they 
are nearly equal and opposite in sign. The nominal errors in the resulting $SuperK_{S}$ 
magnitudes are 2\% or less. Similar procedures were used to compute $Super K_S$ for the
solar-type stars in Tables C1 and C3, except that no extinction corrections were applied.
Although some of the minor offsets in 2MASS have 
been mitigated by our using only Read 1 observations, there is little 
information over the full Read 1 dynamic range on how the photometry behaves 
at this level of accuracy. We therefore place an estimate of 3\% on the net 
errors - half from the rms errors and half from possible systematic ones. 

\subsection{Procedure with IRS Data}

To obtain a high signal-to-noise IRS spectrum of a solar-type star, 
we started with the spectra of the A stars HR 1014, 
2194, 5467, and HD 163466 and of the solar-type stars HD 9826, 10800, 13974, 
39091, 55575, 84737, 86728, 95128, 133002, 136064, 142373, 188376, 196378, 
212330, and 217014 obtained from the Spitzer archive. After standard 
processing with the S13 IRS pipeline to the BCD level, nod pairs were 
subtracted to remove the background and the spectra were extracted with 
SPICE (Spitzer Science Center 2007). We then ratioed the spectra for similar stellar types in various 
combinations to determine the subset that were most closely similar. We also 
gave weight to evidence from the signal strength that the star was 
accurately centered in the spectrograph slit. As a result, we rejected HR 
1014 from further processing because of a number of broad peaks and valleys 
in the ratio. We rejected the short module data for HD 9826, 95128, 188376 
because of pronounced downward slopes with increasing wavelength in the 
ratios, and also HD 13974 and HD 196378 because the S module ratio is low 
relative to the L module data, suggesting a pointing issue with S. We 
rejected the L module data for HD 13974 and 188376 because of slopes and 
curves in the ratios, as well as for HD 9826, 95128, and HD 142373 because 
low values of the ratio relative to that for the S module suggest pointing 
issues. We averaged the accepted spectra (ten for each module) with equal 
weight, and then divided the solar-type average spectrum with the A-star one 
and multiplied the result by the A-star Kurucz model. The resulting spectrum 
had a discontinuity between the short and long IRS modules, which we removed 
by forcing the average value between 12 and 14$\mu $m to equal that between 
15 and 17$\mu $m. We finally smoothed the result with a 7-pixel boxcar
(giving a final spectral resolution of $\sim$ 10\%).

\eject

\section{Appendix D. JHK System}
\subsection{Vega and the Zero Point}
Most near infrared photometric systems claim to be relative to a
zero point defined by Vega, with this star or a model of it placed
either at 0.02 (to coincide with the Johnson bright star measurements)
or at zero. We have investigated how the peculiarities of this star
have affected these definitions. This question is not easily
answered because few of the sets of photometry since Johnson
have actually measured Vega directly. We have therefore evaluated 
the quality of the Johnson photometry and then transformed it
into the 2MASS system, finally using the transformation to compute
the magnitude of Vega as it would have been measured by 2MASS. 

Johnson et al. (1966) report 173 measurements of Vega within 
the overall set of photometry that constitutes their study of
bright stars. We evaluated the quality of this photometry in
two ways. First, we looked at the internal scatter for the 
individual measurements of Vega. We rejected the four highest
and lowest measurements at both J and K (i.e., 4.6\% of
the measurements) and computed the straight
average and standard deviation of the remaining 165 measurements. 
The results are J = 0.019 $\pm$ 0.004, K = 0.023 $\pm$ 0.004,
and rms scatter of 0.048 and 0.050 in J and K respectively. 
With no outlier rejection the values for Vega are unaffected,
but the errors increase to 0.005 and the rms scatter values to
0.061 and 0.065. However, we believe the values with the outlier
rejection are the most representative. 

We have tested this result in another way, by comparing stars measured
in common by Johnson et al. (1966) and either Bouchet et al. (1991) or
Kidger and Mart\'in-Luis (2003). We first transformed the Johnson
photometry into the other system (as described below). We excluded
stars used by Johnson et al. (1966) as standards (stars such as Vega 
that were measured many times) in case their use in fitting for
the photometric corrections would bias the comparison. There were
314 measurements suitable for the comparison with Bouchet et al. 
(1991), of which we rejected
the seven high and seven low outliers (i.e., 4.5\% of the measurements). 
The rms of the deviations was then 0.054 at J and 0.048 at K, that is,
in excellent agreement with the internal scatter of the Johnson
measurements of Vega. Similarly for the measurements of Kidger
and Mart\'in-Luis (2003), there were 135 suitable measurements. 
After transforming the Johnson photometry and rejecting the two
high and two low outliers (3\% of the measurements), the rms scatter
is 0.048 at both J and K. We conclude that the Johnson et al. (1966)
photometry is valid at the level of 1-$\sigma$ errors of 0.05 magnitudes
for single measurements. Since most of the published bright star
photometry is based on three or more measurements per star, this
result is consistent with a typical overall accuracy of 3\% or better. 

The accuracy of the Johnson et al. (1966) photometry implies that
Vega is tied into their overall photometric system to a 1-$\sigma$
error of less than 0.005 magnitudes. We will now determine transformations to
put this measurement into the 2MASS Read 1 system. The simplest approach
is to transfer directly. However, there are relatively few stars that
were measured to high accuracy with the original Johnson photometer
and that are faint enough not to saturate the 2MASS detector arrays. 
We have based the transformation on stars measured in Johnson et al. (1966, 1968) 
and in Lee (1968). The comparison is compromised by the
decreased signal to noise of the stars and we have not attempted
a color correction (although the data imply it would be small).
We derive that Vega would have a 2MASS J magnitude of -0.042 and
a K magnitude of -0.057. Given the issues with signal to noise,
we prefer the average of these values, -0.050, as a best estimate of the
near infrared Vega magnitude.

This result is limited by the poor overlap
in the dynamic range of the two measurement sets; to overcome
this problem, we have calculated the 2MASS Vega magnitudes
by transforming the Johnson photometry to other systems that
have a wider overlap with 2MASS. For the first case, we
used bright standards in the CIT system defined by Elias et al.
(1982). We find a transformation in good agreement with those
derived previously by Elias et al. (1985) and Bessell and Brett (1988),
although our procedure differed from theirs because we fitted the
individual measurements of Johnson et al. (1966) not the final
photometric results combining those measurements for each star. 
We use the transformation from the CIT to 2MASS systems
available at the 2MASS web site to obtain 2MASS K$_S$ magnitudes for
Vega of -0.028, -0.023, and -0.032 using respectively our transformation 
and those of Besell \& Brett (1988) and of Elias et al. (1985). 
Similarly, we obtain a 2MASS K$_S$ magnitude of -0.033 if we make
the transformation through the ESO system photometry of Bouchet et al. (1991)
and a K$_S$ magnitude of -0.045 if the transformation is made through
the main body of photometry of Kidger \& Mart\'in-Luis (2003; their
Table 2). There is a conflict in the latter case because 
Kidger \& Mart\'in-Luis (2003) also measured Vega directly,
using a reduced system gain, and set it identically to zero
magnitude at the near infrared bands (their Table 3). To
check this result, we have computed the average K - L 
(rejecting one high and one low value) for 
the 19 late B stars and A stars using L from their Table 3 (in all cases
measured with the reduced gain) and K from either Table 2 or 3.
We obtain + 0.058 $\pm$ 0.016 where we would have expected
zero. This result implies that the L magnitude of Vega
if measured relative to the high-gain K photometry would
be -0.058 $\pm$ 0.016, in agreement with the negative
K magnitude derived from their Table 2. Although we have been unable
to explain the discrepancy completely, the behavior of
the other stars at L has led us to accept the value from
the transformation of the data in Table 2. An additional comparison
can be made from Koornneef (1983), who established a system 
based on the Johnson J and K photometry. From the 2MASS
color transformations, the corresponding 2MASS K magnitude
for Vega would be -0.023.

All of these estimates are consistent with assigning Vega
the average K magnitude of -0.036 $\pm$ 0.006 in the 2MASS system.
The quoted error is based on the scatter of the measurements; given
the small number of them, an error of 0.010 is a more secure estimate. 

\subsection{2MASS Measurements}
The 2MASS data provide a very homogeneous set of near infrared photometry
over the entire sky, with accuracies well within the original specifications for the survey.
Nonetheless, there are small offsets in those measurements that can be 
significant at the levels of accuracy desired for the calibration discussed
in this paper. We discuss two such effects in this appendix.

\subsubsection{Read1/Read 2 offset}

There are offsets of the order of 2\% 
between Read 1 and Read 2 measurements, in the range where they overlap. 
It is not clear how to derive a universal correction
for this effect. Therefore, for the calibration we selected stars measured
only in Read 1 mode.  

\subsubsection{Procedures for VJHK Calibration of A Stars}

To define a zero point at V, J, H, and K, we built a 
sample of 57 A0V stars distributed over the entire 
sky, brighter than m$_{V}$ = 7 mag., and with "A" quality Read 1 2MASS 
measurements at all three near infrared colors. We took B and V magnitudes from the 
Hipparcos Main Catalog (Perryman et al. 1997). Stellar classifications were 
taken from SIMBAD. The infrared colors are not a strong function of the 
spectral type - from B8 to A2, J - K changes from -0.03 to +0.03 (Tokunaga 
2000) - so type errors are not a major concern. However, reddening is. To 
eliminate reddened stars (and other sources of error), we excluded ones with 
B-V $>$ 0.05 and ones with $|$V - K$_S$$|$ $>$ 0.05. In 
addition, because the 2MASS quoted errors can be large in H band, we 
excluded three stars with H - K $>$ 0.07, which is 5\% redder than the value 
averaged over the sample. For the final sample of 57 stars (see Table D1), 
the result was an average $<$V - K$>$ consistent with zero, $<$J - K$ >$ = -0.022 
$\pm$ 0.003, and $<$H - K$>$ = 0.019 $\pm$ 0.003. The average 
values are not affected by various assumptions we tried with regard to final 
sample selection. For example, if we keep all stars with $|$V - K$|$ $<$ 0.1, 
we get a slight positive residual in $<$V - K$>$, but values for $<$J - K$>$ 
and $<$H - K$>$ equal to the ones above within the errors. 

An independent test was conducted with A0 dwarfs from the Michigan spectral atlas. 
There are a total of 6008 such stars with "good" 2MASS measurements in all 
three bands. We discarded all of them with $|$V - K$_S$$|$ $>$ 0.05, leading 
to a sample of 510. Eleven of this sample were discarded as photometric outliers. 
Of the remaining stars, 220 have Read 1 photometry in all three bands, and the 
net J - K$_S$ = -0.016 $\pm$ 0.018 for these stars. Similarly, we found a net H - K$_S$ = +0.026 $\pm$ 0.022 
based on 212 stars from the Michigan spectral atlas sample. The quoted errors are the population 
standard deviations. The nominal standard deviations of the means are 
only about 0.001, but there are suggestions of other low level trends in the 
data that would make such an accurate color difficult to obtain. We conclude 
that the net colors are in agreement with the ones derived 
from the smaller and more heterogeneous sample discussed above. The result for the stars 
measured in Read 2 mode are J - K$_S$ = -0.005 $\pm$ 0.021 (population sigma, 
150 stars) showing that the offset is a function of the 2MASS read mode. Therefore, 
corrections to remove these small effects will be relatively complex to apply.
Cohen et al. (2003) have also identified offsets of similar size, but not
with the same values as those we have determined. Their sample is predominantly
K giants, and their Vega zero point model is redder than ours. If we compute the 
offsets only for the nine A stars in their sample and correct for the different
colors for Vega, their results agree with ours to within 1.5 - 2\% with errors of
about 1.6\%, i.e., within the errors. 

These results could at least partially explain infrared color offsets such as those 
found by Casagrande et al. (2006; Figures 2,3, and 4). Future improvements in the
comparisons with theoretical models will need to take account carefully of such small
systematic effects in the photometry. 

\subsubsection{Procedures for Solar-Type Stars}

In the V, J, H, and K bands, the colors of solar-type stars depart 
significantly from a Rayleigh-Jeans shape and we have to be careful 
in defining a sample of such stars. We determined average 
colors from the list of 36 solar-type stars in Table D2. The stellar 
parameters are from the Appalachian State Nstars web site (Gray 2007), with 
a few extras from Soubiran \& Triaud (2004). A fit of temperature vs. 
spectral type from the full Nstars database shows that the average G2V star 
is assigned a temperature of 5720K. The solar temperature of 5778K is within 
the scatter, but either the sun is relatively hot for its spectral class, or 
there is a systematic offset between its temperature and those assigned 
through the stellar models to other stars. 

We used the Hipparcos 
photometry at V to eliminate stars from the sample with V - K$_{S}$ colors 
differing by more than 0.1 magnitudes from expectations for their spectral 
types. These stars are systematically much closer than the A stars discussed 
above, so the color deviations are less likely to arise from 
reddening. In support of this conclusion, there are nearly as many blue as 
red deviations. We attribute the discordant values to errors in the spectral 
types.

Each V - K$_{S}$ was corrected to the equivalent value for 5778K according 
to a relationship fitting the temperature of the star and the standard V - K 
colors from Tokunaga (2000). This relationship was set to zero at 5778K, so 
it acts only to take out the variations with temperature but not to cause an 
overall shift in color. We ignored reddening because the maximum distance of 
the sample members, 40pc, puts them within the Local Bubble of very low 
density interstellar medium (e.g., Knude \& Hog 1998 and references 
therein). 

Although we prefer to use the Hipparcos V magnitudes because they are 
homogeneous over the entire sky, to do so we need to allow for possible 
systematic offsets relative to Johnson V. These effects should be small for 
A stars because of their nearly zero colors. To evaluate them for solar-type 
stars, we used the sample of 102 stars listed in the Nstars database as 
being of G-type and for which there was high quality V photometry, to find a 
net offset of $<$V$_{Hipp}$ - V$_{Johnson}$$>$ = -0.007 $\pm$ 0.002. 
We corrected the measured V - K$_{S}$ color accordingly. We also found two 
stars in our solar-type sample, HD 139777 and HD 218739, with large offsets 
between the Nstars tabulated groundbased magnitudes and those from Hipparcos 
(0.07 - 0.08 mag.). Multiple groundbased measurements plus the V - K color 
showed that the groundbased results were correct, and we substituted them. 
Otherwise, the standard deviation in the difference of Hipparcos and 
groundbased V magnitude derived from uvby was only 1.6\%, confirming the 
high quality of the Hipparcos data (and of the groundbased uvby data also).

\eject

\section{Appendix E. Calculation of Effective, Nominal, and Isophotal Wavelengths}

Given \textit{R($\lambda $)} as 
the\textit{} relative response function of the photometric band, the 
equivalent band width is

\begin{equation}
BW = \frac{{1}}{{R_{max}} }\int {R\left( {\lambda}  \right)\;d\lambda } 
\end{equation}

\noindent
where \textit{R}$_{max}$ is the maximum value of \textit{R($\lambda $)} 
(often \textit{R($\lambda $)} is normalized to 1). The average flux in the 
band is

\begin{equation}
 \left\langle {F_{\lambda} }  \right\rangle = \frac{{\int {\lambda 
\;F_{\lambda}  \left( {\lambda}  \right)\;R\left( {\lambda}  
\right)\;d\lambda} } }{{\int {\lambda \;R\left( {\lambda}  \right)\;d\lambda 
}} }\;,
\end{equation}

\noindent
or equivalently is the total flux 
passed by the band divided by the band width.  

To idealize measurements made through a given photometric band to a
monochromatic equivalent flux, we
need to assign a specific wavelength to the measurement. 
The mean, or effective, wavelength is defined by the following equation:

\begin{equation}
\lambda _{0} = \frac{{\int {\lambda} \, R \left( \lambda  \right) \;d\lambda 
}}{{\int R\left( {\lambda}  \right) \;d\lambda  }}.
\end{equation}

\noindent
Examples are given in Table 6. The zero point (ZP) is conventionally defined as 
the flux density of "Vega" at $\lambda_0$, but the stellar absorption
features pose complications. The ZPs in Table 6 are determined by
interpolating over the absorptions in the stellar spectrum and then 
averaging $\lambda^4 F_\lambda$ over a 1\% bandpass. 
Any measurement expressed in terms of the mean wavelength of the photometric 
band needs to be accompanied by a color correction term that converts the 
apparent flux density to the value at $\lambda _{0}$ for the specific 
source spectral energy distribution that reproduces the observed signal 
strength (e.g., Low \& Rieke 1974). The color correction is normalized to 
the result for a source with \textit{F}$_{\lambda} $\textit{($\lambda $)} = 
constant across the band. The absorption features in stellar spectra can make
computation of a precise color correction problematical, but alternatively
one can be based on a blackbody spectrum. We have taken this approach
for the illustrative values in Table 6. They are given in the sense that
a flat spectrum that gives the same signal as a blackbody 
of the specified temperature will be 
brighter at the mean wavelength by the tabulated factor. It is convenient in many cases to set the 
color correction to zero for hot stars, which can be accomplished by an offset in 
the set of color terms with an appropriate counter-adjustment in the nominal 
zero point flux density (e.g., Low and Rieke 1974; MIPS Handbook).

Another way to circumvent the effects of stellar absorptions is to use
the "average zero point", $\left\langle {F_{\lambda} }  \right\rangle$, where the
flux density is that of "Vega".  
Where the stellar spectrum is smooth and follows a Planck curve closely,
the average flux in the band is reproduced reasonably well 
by the stellar flux density at the mean wavelength times
the color correction. Where the stellar spectrum deviates from Planckian
behavior (e.g., in the H band due to the many members of the Brackett series), 
there may be significant deviations from this relation. In general cases,
the use of a fiducial wavelength and zero point imposes limits in the achievable
accuracy; it may be required to carry out the relevant integrals of the relative
response function convolved with the object spectra. 

Under many conditions, the value of the required color correction can be 
minimized by modifying the wavelength associated with the band to the 
"nominal wavelength" (Reach et al. 2005), defined as

\begin{equation}
\lambda _{0} ^{\prime}  = \frac{{\int {\lambda ^{2}\;R\left( {\lambda}  
\right)\;d\lambda} } }{{\int {\lambda \;R\left( {\lambda}  \right)\;d\lambda 
}} }\,.
\end{equation}

\noindent
Table 6 includes the ZP flux densities appropriate to this definition.
Color corrections under this definition are relative to those for a source 
with \textit{F}$_{\lambda} $\textit{($\lambda $)} $ \propto \lambda $. 
Otherwise, the procedures and pitfalls are similar to those for the mean wavelength.

The definitions in equations (3) and (4) have the advantages that they are 
simple in concept and that the wavelength associated with a band is 
independent of the source properties. Yet another definition is the 
isophotal wavelength, defined by

\begin{equation}
F_{\lambda}  \left( {\lambda _{iso}}  \right) = \left\langle {F_{\lambda}. }  
\right\rangle 
\end{equation}

\noindent
This approach has the advantage of subsuming the correction terms into the 
wavelength specification, providing a more streamlined description of a 
measurement. However, it is potentially confusing to have different 
wavelengths associated with measurements of different sources in the 
identical photometric band, and relating measurements to the identical 
wavelength requires introduction of color corrections similar to those 
needed with the definitions in equations (3) and (4) (Tokunaga \& Vacca 
2005). 

The issue of stellar features is complex in the use of $\lambda _{iso}$, where the basic 
definitions can be misleading (e.g., \textit{F}$_{\lambda 
}$\textit{($\lambda $}$_{iso}$\textit{)} may depend on the resolution used 
to measure the spectrum of the source). To avoid this difficulty, one must use
a continuum model of the stellar spectrum or interpolate over its
spectral features. Golay (1974) suggests avoiding these 
difficulties by replacing the stellar spectrum with a suitable Planck 
function. Instead, for the 2MASS bands we have interpolated over any lines 
close to the fiducial wavelength (a procedure similar to that used by Cohen 
et al. 1992, 2003). We find that reliable interpolations can be based on 
\textit{$\lambda $}$^{4}${\textit{ F}$_{\lambda}$}. Since the 
infrared is close to the Rayleigh-Jeans realm for Vega, the resulting 
function varies slowly and smoothly with wavelength, making it appropriate 
to use linear interpolations across line-contaminated regions. The interpolated 
values were converted to flux by dividing by \textit{$\lambda $}$^{4}$. 
Figures E1 - E3 show our interpolations on the model spectrum of "Vega". 
Table E1 lists the beginning and end wavelengths for the interpolations. Our 
calculated values of \textit{$\lambda $}$_{iso}$ in Table 6 agree closely 
with those of Cohen et al. (2003). Therefore, many aspects of their 
analysis, such as the water vapor dependence of the nominal bands and 
fiducial wavelengths, can be applied to our proposed calibration without 
change.

For the mid-infrared space missions, color corrections as a function of 
source spectral characteristics can be found in the appropriate user 
handbooks. The MIPS calibration is to a mean wavelength, while the IRAC and 
IRAS ones are to a nominal wavelength. Because these corrections are tracked 
as official project values, we quote our results in a form that assumes they 
will be used as tabulated.

\clearpage

\begin{figure}
\plotone{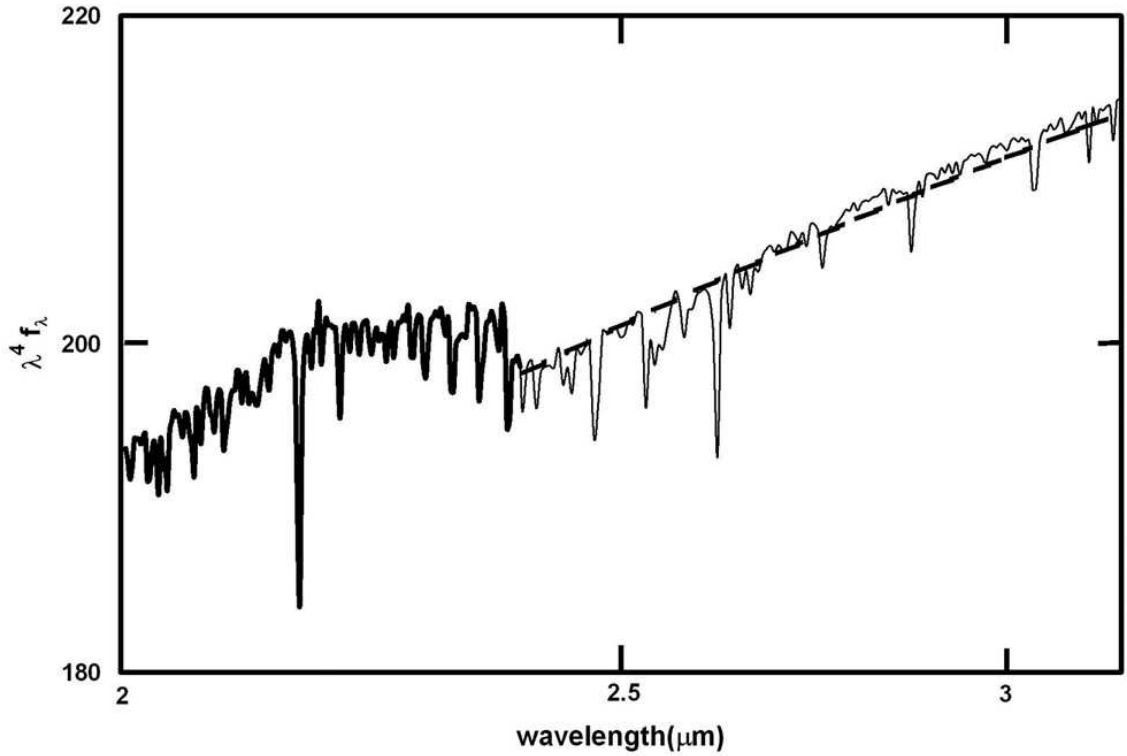}
\caption{\label{calfig0}
Joining the segments for a complete solar spectrum. The heavy line is the measured spectrum
from Thuillier et al. (2003). The dashed heavy line is the Engelke approximation. Both
of these curves are left in the measured units with no re-normalization. The light solid
line is the modified HM74 model that we use to represent the empirical solar spectrum,
normalized to provide a smooth transition from the Thuillier et al. spectrum consistent
with the Engelke approximation.}
\end{figure}

\clearpage
\begin{figure}
\plotone{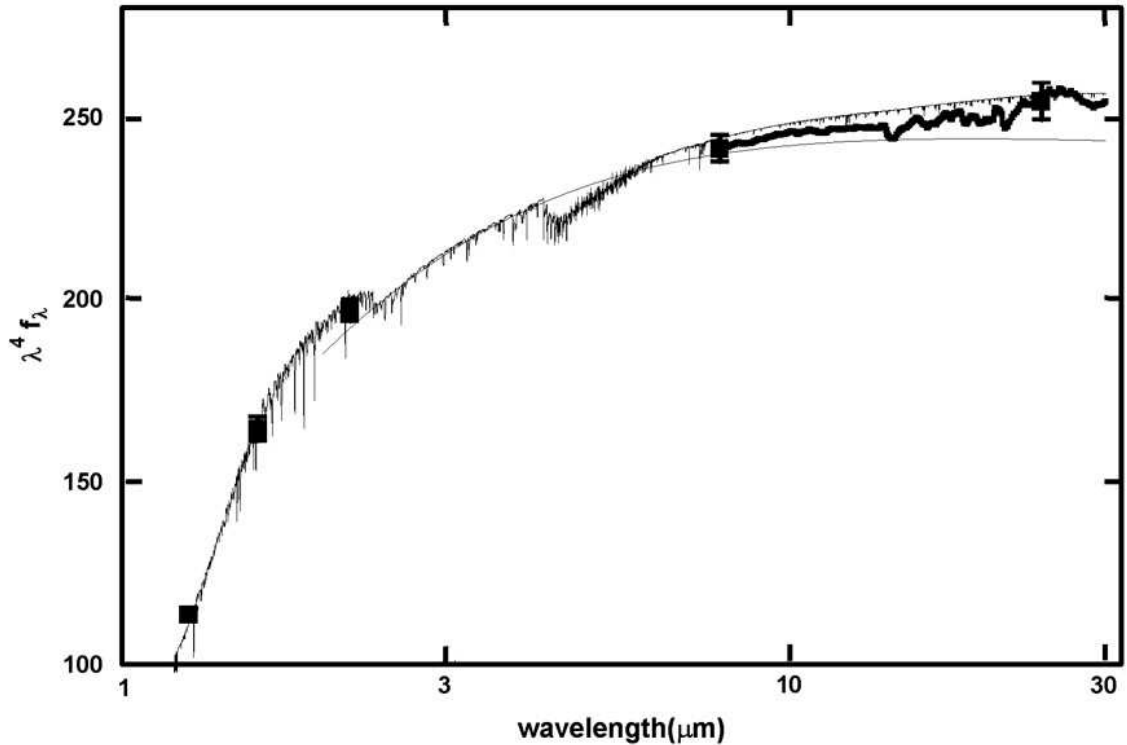}
\caption{\label{calfig1}
Infrared spectrum of the sun. The spectrum has been multiplied by $\lambda^4$ to 
facilitate detailed comparisons in the infrared. Out to 2.4$\mu$m, we plot the 
measured solar spectrum, based on the results of Thuillier et al. (2003). 
Beyond this wavelength, the solar spectrum is represented by the HM74 model,
normalized and modified at the CO fundamental bands as described in the text. The photometry 
of solar type stars (normalized at V) is shown as 
square points, and we have put a 2\% error bar on the measurement at 24$\mu$m. Between 8 
and 32$\mu$m, we also show the spectrum of solar-type stars obtained with the Spitzer 
Infrared Spectrograph (black curve). The smooth curve below the spectrum is the
Engelke (1992) approximation to the solar continuum.}
\end{figure}

\clearpage

\begin{figure}
\figurenum{B1}
\plotone{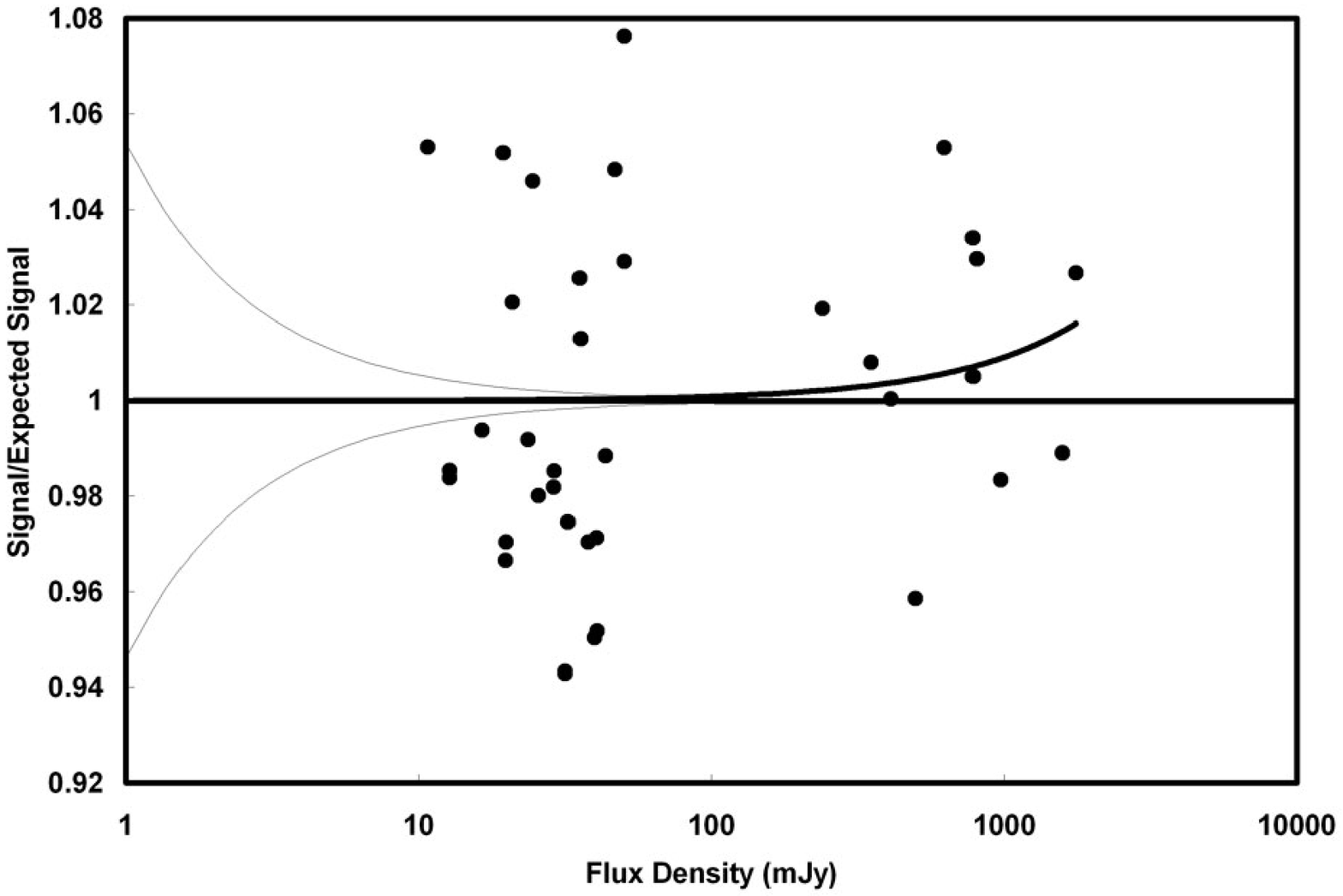}
\caption{\label{calfig2}
Normalized (to linear response) Signal vs. Flux Density at 24$\mu$m. 
Models for three different types of nonlinearity are compared with the data. The two light 
lines are nonlinearity due to residual positive or negative 50$\mu$Jy latent images. The 
heavy line is a linearity correction; the best fit suggests that the correction applied 
in the data pipeline may be slightly too large (but not at a significant level).  However, 
there is no convincing argument for any significant nonlinearity over the range of the calibration 
measurements (10 - 50mJy).}
\end{figure}

\clearpage

\begin{figure}
\figurenum{E1}
\plotone{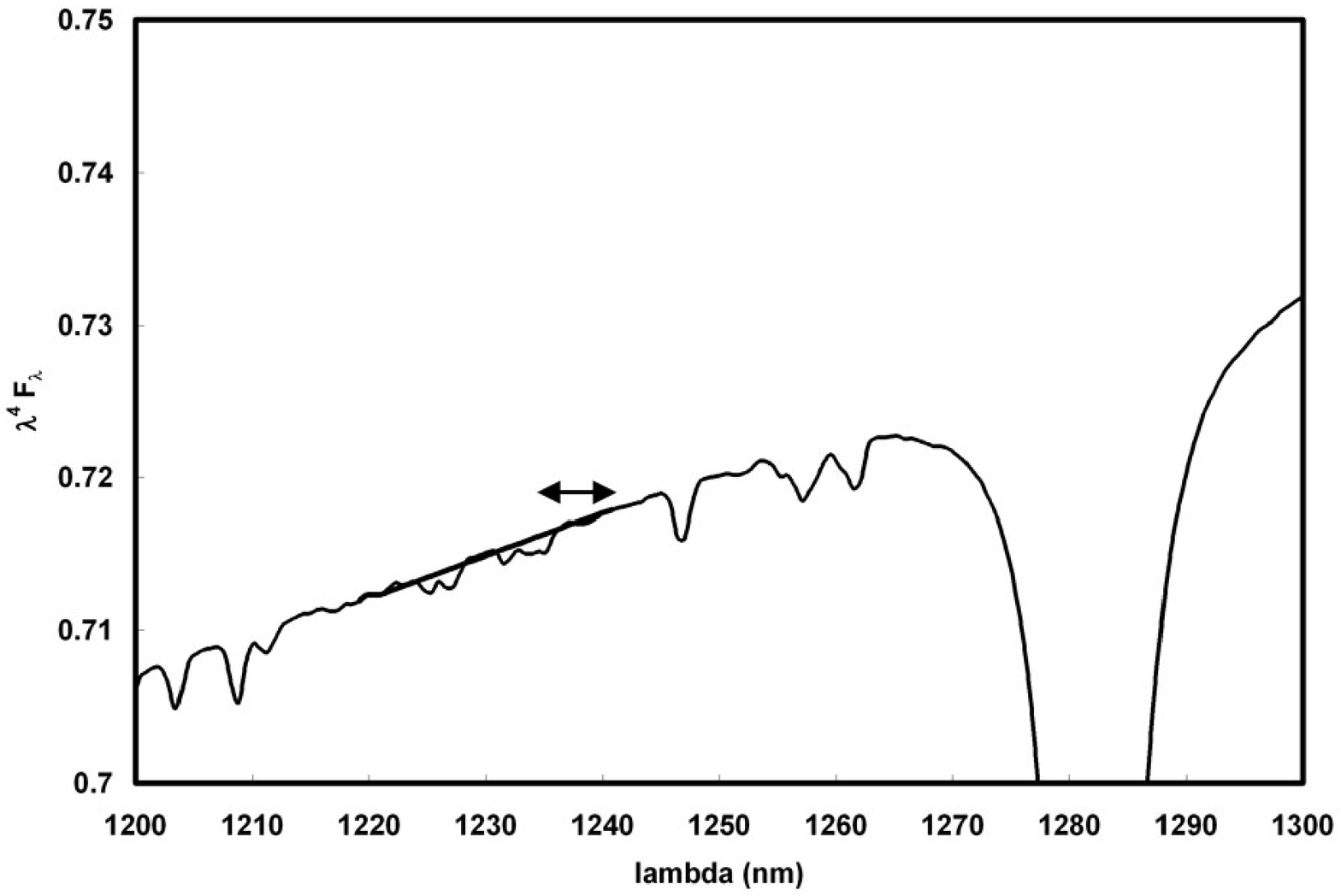}
\caption{\label{calfig3}
Interpolation over spectral absorptions in the J band. The horizontal arrow shows 
the approximate range of the various band-defining wavelengths.}
\end{figure}

\clearpage
\begin{figure}
\figurenum{E2}
\plotone{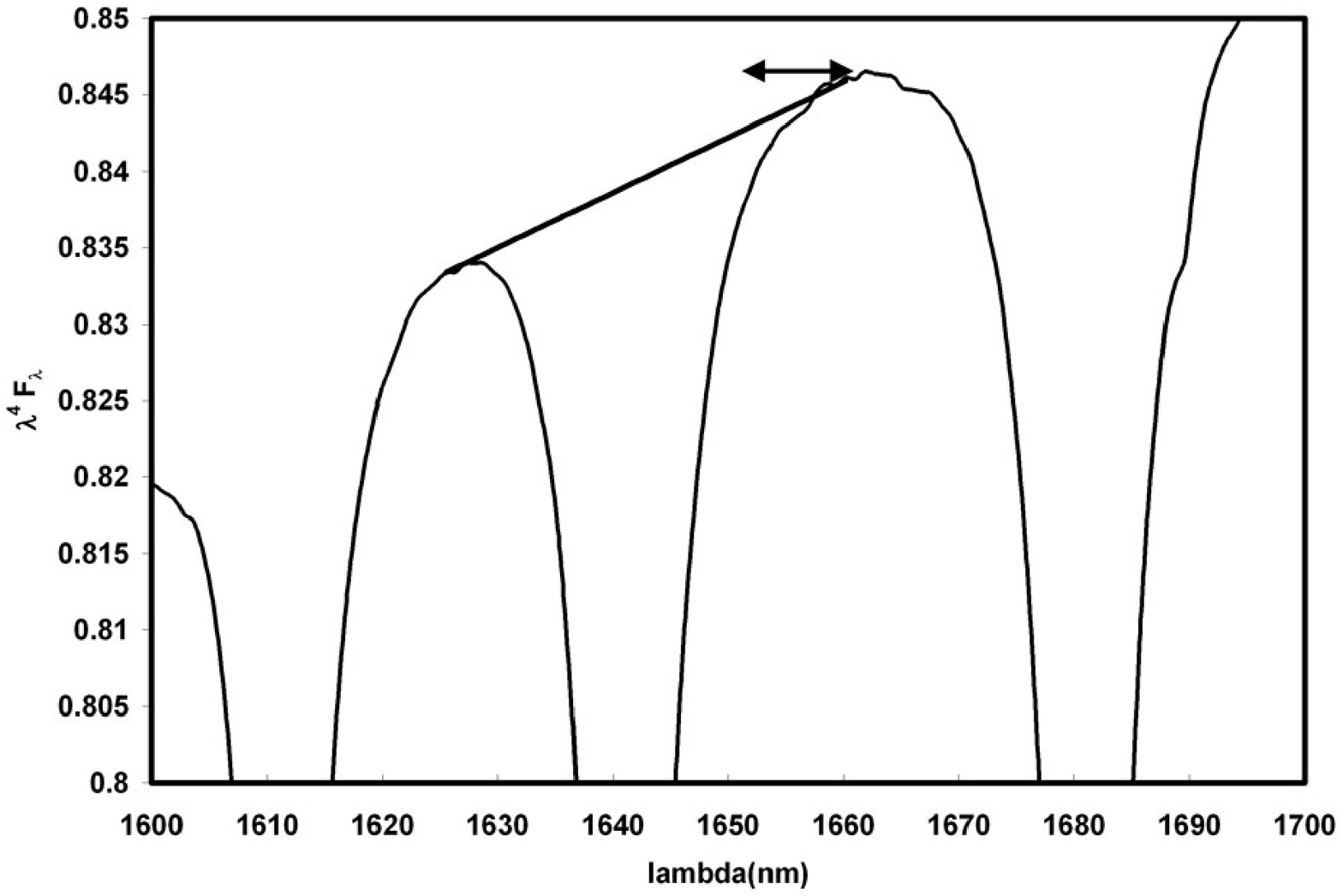}
\caption{\label{calfig4}
Interpolation over spectral absorptions in the H band. The horizontal arrow shows 
the approximate range of the various band-defining wavelengths.}
\end{figure}

\clearpage
\begin{figure}
\figurenum{E3}
\plotone{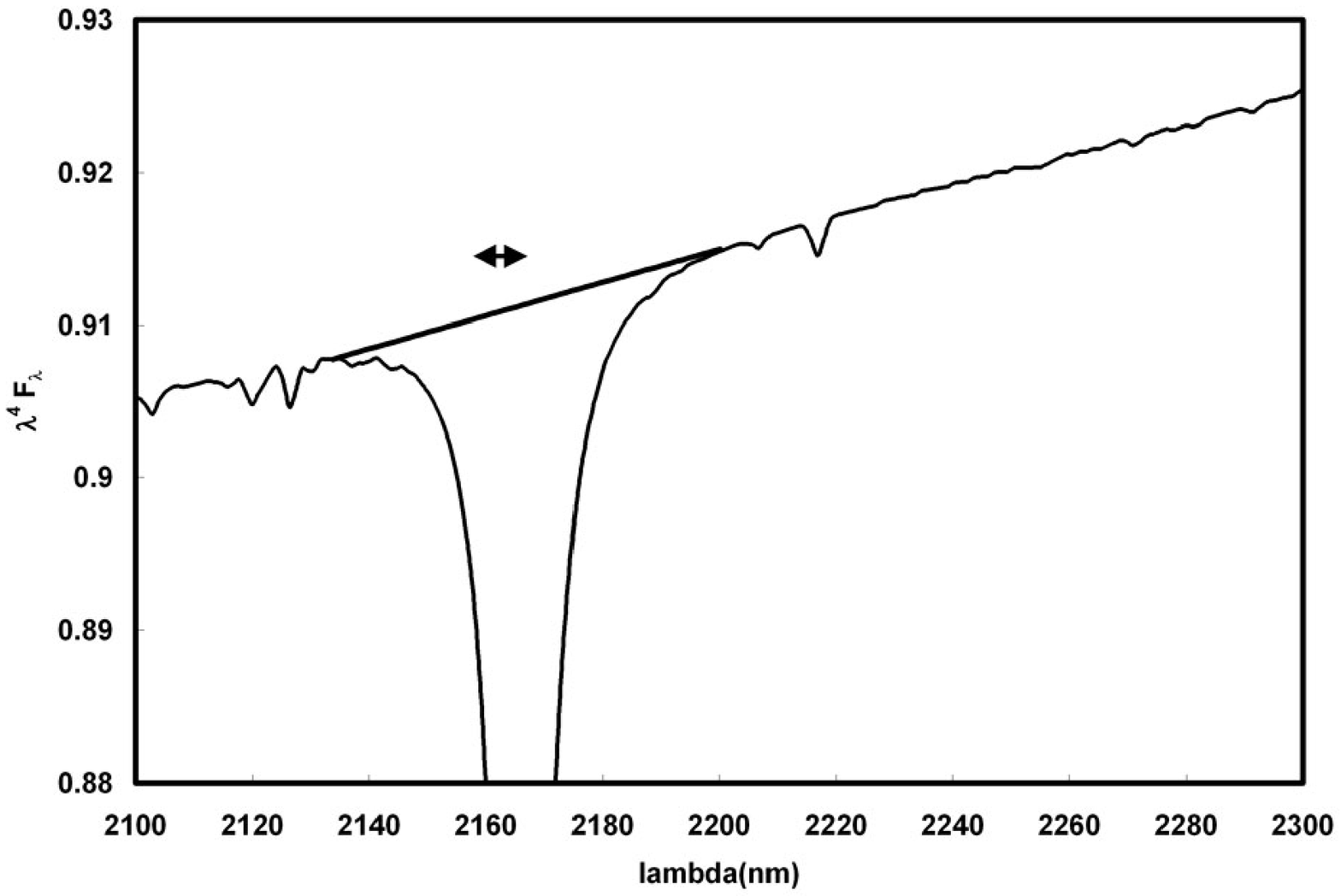}
\caption{\label{calfig5}
Interpolation over spectral absorptions in the K$_S$ band. The horizontal arrow 
shows the approximate range of the various band-defining wavelengths.}
\end{figure}

\clearpage

\begin{deluxetable}{lll}
\tablecolumns{3}
\tabletypesize{\scriptsize}
\tablecaption{ \label{10micron} Absolute Flux Density from "Vega" at 10.6 and 24$\mu $m}
\tablewidth{0pc}
\tablehead{\colhead{Approach}  &  \colhead{Flux Density }  &  \colhead{Error } \\
\colhead{}  &  \colhead{@ 10.6$\mu$m (Jy)}  &  \colhead{(Jy)}
}
\startdata
Rieke et al. (1985) & 35.3 & 1.1 \\
MSX weighted average & 35.04 & 0.24 \\
Weighted average of Rieke, MSX & 35.05 & 0.23 \\
Solar analog (this work) & 34.53 & 1.1 \\
All measurements weighted average & 35.03 & 0.23 \\
Average Rieke et al., solar analog & 34.92 & 0.8 \\
\textbf{Adopted} & \textbf{35.03} & \textbf{0.3} \\
Adopted, F$_\lambda$ & 9.35X10$^{-17}$ W cm$^{-2}$ $\mu$m$^{-1}$ &  \\
Hammersley et al. 1998 & 35.2  &  --  \\
Flux Density at 23.675$\mu$m & 7.17  &  0.11  \\

\enddata
\end{deluxetable}

\begin{deluxetable}{lll}
\tablecolumns{3}
\tabletypesize{\scriptsize}
\tablecaption{ \label{2micron} Measurements of the Absolute Flux of "Vega" at 2.22$\mu $m}
\tablewidth{0pc}
\tablehead{\colhead{Reference}  &  \colhead{Equivalent 2.22$\mu$m }  &  \colhead{Error } \\
\colhead{}  &  \colhead{Flux Density (Jy)}  &  \colhead{(Jy)}
}
\startdata
Walker (1969) & 638 & 64 \\
Blackwell et al. (1983) & 666 & 20 \\
Selby et al. (1983) & 623 & 25 \\
Booth et al. (1989) & 667 & 27 \\
Weighted average & 653 & 13 \\
Corrected for disk & 645 & 15 \\
Extrapolated from 10.6$\mu $m & 649 & 10 \\
\textbf{"best" calibration} & \textbf{647} & \textbf{8} \\
\enddata
\end{deluxetable}

\begin{deluxetable}{lllll}
\tablecolumns{5}
\tabletypesize{\scriptsize}
\tablecaption{ \label{comparison} Comparison of Solar and Stellar Colors}
\tablewidth{0pc}
\tablehead{\colhead{Band}  &  \colhead{Synthetic Solar} &  \colhead{Observed, }  &
 \colhead{HM74 }   &   \colhead{  Fontenla et al. }  \\
\colhead{}  &  \colhead{Color (mag)} &  \colhead{Solar-Type Stars}  &
 \colhead{Model}   &   \colhead{2006 Model} 
}
\startdata
V$_{J}$ & 0.00   & 0.00 &   & 0.00 \\
V-J       & 1.158 $\pm$ 0.02 & 1.158 $\pm$ 0.015 &   & 1.20 \\
V-H       & 1.513 $\pm$ 0.02 & 1.484 $\pm$ 0.020 &   & 1.55 \\
V-K$_{S}$ & 1.568 $\pm$ 0.02 & 1.545 $\pm$ 0.015 & 1.550 & 1.57 \\
V-$\rm{[8]}$ & 1.596 $\pm$ 0.02 & 1.591 $\pm$ 0.015 & 1.596* & 1.615 \\
V-$\rm{[24]}$& 1.54  $\pm$ 0.05 & 1.590 $\pm$ 0.020 & 1.577  & 1.564 \\
\enddata
\tablecomments{*Adopted value, since model does not extend below 2$\mu $m; the 
Thuillier et al. SED was used to fill in the 1.927 - 2.00$\mu $m range not 
included in the model. }
\end{deluxetable}

\begin{deluxetable}{lll}
\tablecolumns{3}
\tabletypesize{\scriptsize}
\tablecaption{ \label{bessell} Solar analog colors corrected to our system from Bessell et al. (1998)}
\tablewidth{0pc}
\tablehead{\colhead{}  &  \colhead{$V-K_S$}  &  \colhead{Reference} \\
}
\startdata
Sun-ref    &  1.574              & Colina et al. 1996               \\
Analog       &  1.563              & Cayrel de Strobel 1996, Table 6  \\
Model        &  1.587              & SUN-OVER*                         \\
Model        &  1.587              & SUN-NOVER*                        \\
\hline
             &  from Table 3           &        \\
Solar Spectrum &  1.568$\pm$ 0.02      & V, K$_S$ from Thuillier et al. 2003 \\
Model         &  1.57                  & Fontenla et al. 2006        \\
ATLAS9 Model   &  1.556                & Casagrande et al. 2006      \\
Kurucz 2004    &  1.555                & Casagrande et al. 2006      \\
MARCS Model    &  1.547                & Casagrande et al. 2006      \\
Solar-Type Stars &  1.545$\pm$0.02     & This work     \\
\hline
         &  NICMOS prime calibrator    &         \\
P330E     &  1.577                     & Bohlin et al. 2001             \\

\enddata

\tablecomments{*Terminology from Bessel et al. (1998); SUN-OVER refers to 
ATLAS9 models with overshooting turned on, while SUN-NOVER has it turned off. }

\end{deluxetable}

\begin{deluxetable}{llll}
\tablecolumns{4}
\tabletypesize{\scriptsize}
\tablecaption{ \label{midirfact} Multiplicative Factors to Reconcile Mid-IR Calibrations to Proposed One}
\tablewidth{0pc}
\tablehead{\colhead{2MASS K$_S$} & \colhead{IRAC \par 7.872$\mu $m}  &  \colhead{IRAS 12$\mu $m }  &  \colhead{IRAS 25$\mu $m} \\
}
\startdata
1.02* & 1.015**  & 0.992  & 0.980 \\

\enddata
\tablecomments{*Relative to calibration of Cohen et al. (2003). **Relative to calibration of Reach et al. (2005)}
\end{deluxetable}

\begin{deluxetable}{llll}
\tablecolumns{4}
\tabletypesize{\scriptsize}
\tablecaption{ \label{2masscal} Suggested calibration of 2MASS photometry
}
\tablewidth{0pc}
\tablehead{\colhead{}  &  \colhead{}  &  \colhead{}  &  \colhead{} \\
}
\startdata
Band & J  &  &  \\
Band Width  & 0.1625$\mu $m &  &  \\
Average Zero Point (ZP) (W cm$^{-2}$ $\mu$m$^{-1}$) & 3.21 X 10$^{-13}$ &  & \\
Wavelength Type  & $\lambda _{0}$ ($\mu $m) & $\lambda _{0}$' ($\mu $m) & $\lambda _{iso}$ ($\mu $m) \\
Wavelength  & 1.2410  & 1.2444  & 1.2356 \\
ZP F$_\lambda$ at $\lambda_{0}$ (W cm$^{-2}$ $\mu$m$^{-1}$) & 3.16X10$^{-13}$ & 3.13X10$^{-13}$ & 3.21X10$^{-13}$ \\
ZP F$_\nu$ at fiducial $\lambda$ (Jy) &  1623 &  1617 & 1635 \\
Color Correction/9550K Black Body & 1.017 & 1.028 &  \\
\hline
Band & H  &  &  \\
Band Width & 0.2508$\mu $m  &  &  \\
Average ZP (W cm$^{-2}$ $\mu$m$^{-1}$)  & 1.164 X 10$^{-13}$  &  &  \\
Wavelength Type  & $\lambda _{0}$ ($\mu $m) & $\lambda _{0}$' ($\mu $m) & $\lambda _{iso}$ ($\mu $m) \\
Wavelength & 1.6513  & 1.6551 & 1.6597 \\
ZP F$_\lambda$ at fiducial $\lambda$ (W cm$^{-2}$ $\mu$m$^{-1}$) & 1.182X10$^{-13}$ & 1.174X10$^{-13}$ &  1.163X10$^{-13}$ \\
ZP F$_\nu$ at fiducial $\lambda$ (Jy) &  1075 &  1073 & 1068 \\
Color Correction/9550K Black Body  & 1.014 & 1.023 &  \\
\hline
Band & K$_{S}$  &  &  \\
Band Width  & 0.2620$\mu $m  &  &  \\
Average ZP (W cm$^{-2}$ $\mu$m$^{-1}$) & 4.37X10$^{-14}$  &  &  \\
Wavelength Type & $\lambda _{0}$ ($\mu $m) & $\lambda _{0}$' ($\mu $m) & $\lambda _{iso}$ ($\mu $m) \\
Wavelength & 2.1657 & 2.1692 & 2.1598 \\
ZP F$_\lambda$ at fiducial $\lambda$ (W cm$^{-2}$ $\mu$m$^{-1}$) & 4.32X10$^{-14}$ & 4.30X10$^{-14}$ &  4.36X10$^{-14}$\\
ZP F$_\nu$ at fiducial $\lambda$ (Jy) &  676 &  675 & 678 \\
Color Correction/9550K Black Body & 1.018 & 1.030 &  \\

\enddata
\end{deluxetable}

\begin{deluxetable}{lll}
\tablenum{A1}
\tablecolumns{7}
\tabletypesize{\scriptsize}
\tablecaption{ \label{SEDs} Reference Spectral Energy Distributions of
the Sun and Vega}
\tablewidth{0pc}
\tablehead{\colhead{Wavelength}  &  \colhead{Sun}  &  \colhead{Vega} \\ 
\colhead{$\mu$m}  & \colhead{W m$^{-2}$ nm$^{-1}$}  &  \colhead{ W m$^{-2}$ nm$^{-1}$}  \\ 
}
\startdata
0.1998   &   $7.520 \times 10^{-3}$  &  $5.581 \times 10^{-11}$  \\ 
0.2017   &   $8.160 \times 10^{-3}$  &  $5.312 \times 10^{-11}$  \\
0.2035   &   $9.120 \times 10^{-3}$  &  $5.165 \times 10^{-11}$  \\ 
0.2052   &   $1.077 \times 10^{-2}$  &  $5.752 \times 10^{-11}$  \\  

\enddata
\end{deluxetable}

%

\begin{deluxetable}{llllll}
\tablenum{C1}
\tablecolumns{6}
\tabletypesize{\scriptsize}
\tablecaption{ \label{Gstars} Solar Type 8$\mu$m Calibration Sample.}
\tablewidth{0pc}
\tablehead{\colhead{Name}  &  \colhead{Type} & 
\colhead{$SuperK_S$}  & \colhead{K$_S$}  &  \colhead{f(8$\mu$m)}  &  
\colhead{flux ratio*} \\
\colhead{}  &  \colhead{ }  &  \colhead{(mag)}  &  
\colhead{(mag)}  &  \colhead{(mJy)}  &  \colhead{}  \\
}
\startdata
HD 00643&       G2/G3V  &          6.204&          6.187&          215.1&           0.99\\
HD 00894&       F8IV-V  &          5.353&          5.378&          448.8&           0.98\\
HD 01901&       F8      &          5.563&           5.57&          371.6&           0.97\\
HD 02746&       G5      &          6.262&          6.282&          198.5&           1.00\\
HD 03796&       G4V     &          6.279&          6.268&          198.2&           0.98\\
HD 03894&       G1V     &          6.419&          6.442&          172.3&           1.00\\
HD 06073&       G0      &          6.246&          6.253&          202.1&           0.99\\
HD 08820&       G0IV-V  &          6.114&          6.118&          229.9&           0.99\\
HD 08874&       G1V     &          4.967&          4.955&          663.2&           0.98\\
HD 09071&       G2/G3V  &          6.054&          6.042&          235.4&           0.95\\
HD 09278&       G5      &          5.724&          5.721&          322.7&           0.96\\
HD 09855&       G2/G3V  &          6.072&          6.078&          238.8&           0.99\\
HD 10195&       G0V     &          5.611&          5.609&          361.3&           0.97\\
HD 10625&       G0V     &          5.940&          5.929&          270.9&           0.98\\
HD 10879&       G1/G2V  &          4.857&          4.832&          767.1&           1.01\\
HD 10894&       F8      &          6.009&          6.025&          256.9&           1.02\\
HD 11219&       F8V     &          5.556&          5.553&          395.6&           1.01\\
HD 11504&       G1Va    &          5.331&          5.334&          476.0&           1.00\\
HD 12150&       G2V     &          6.135&          6.124&          225.3&           0.98\\
HD 12265&       F8      &          5.863&          5.879&          283.1&           0.98\\
HD 14193&       G2/G3V  &          5.771&          5.764&          310.8&           0.97\\
HD 15070&       G0      &          5.572&          5.565&          388.1&           1.01\\
HD 15922&       G5V     &          4.981&          4.998&          613.3&           0.94\\
HD 17994&       F8V     &          4.952&          4.936&          658.4&           0.96\\
HD 18321&       G2V     &          5.754&          5.766&          320.8&           1.00\\
HD 19301&       F8      &          5.945&          5.966&          264.7&           0.99\\
HD 19503&       G5      &          5.561&          5.579&          379.4&           1.00\\
HD 19959&       G0V     &          5.462&          5.461&          412.2&           0.97\\
HD 20427&       F8V     &          5.412&          5.449&          415.8&           0.97\\
HD 20590&       G4IV-V  &          5.314&          5.321&          491.1&           1.02\\
HD 21229&       G5      &          6.265&          6.269&          206.0&           1.02\\
HD 21627&       G0      &          5.782&          5.794&          309.5&           0.99\\

\enddata

\tablecomments{*Ratio of observed flux density to expected flux density from the photosphere. 
See Rieke et al. (2005) and Su et al. (2006) for further details.}

\end{deluxetable}

\begin{deluxetable}{lllllll}
\tablenum{C2}
\tablecolumns{7}
\tabletypesize{\scriptsize}
\tablecaption{ \label{Astars} A-Type Stars Used to Determine the K$_S$ - [24] Zero Point.}
\tablewidth{0pc}
\tablehead{\colhead{Name}  &  \colhead{Age}  &  \colhead{V } & 
\colhead{$SuperK_S$}  & \colhead{K$_S$}  &  \colhead{f(24$\mu$m)}  &  
\colhead{flux ratio*}
\\
\colhead{}  &  \colhead{(Myr)}  &  \colhead{(mag)}  &  \colhead{(mag)}  &  
\colhead{(mag)}  &  \colhead{(mJy)}  &  \colhead{}  \\
}
\startdata
HD    319  &            600&       A1V     &          5.467&          5.479&          45.89&           0.99\\
HD  02811&            750&       A3V     &          7.067&          7.057&          10.94&           1.01\\
HD  11413&            600&       A1V     &          5.396&          5.422&          53.13&           1.08**\\
HD  14943&            850&       A5V     &          5.469&          5.439&          48.92&           1.01\\
HD  15646&            260&       A0V     &          6.398&          6.411&          20.15&           1.02\\
HD  17254&            650&       A2V     &          5.907&          5.877&          31.12&           0.97\\
HD  17254&            650&       A2V     &          5.907&          5.877&          30.47&           0.94\\
HD  20888&            300&       A3V     &          5.723&          5.691&          37.02&           0.97\\
HD  20888&            300&       A3V     &          5.744&          5.691&          38.28&           1.00\\
HD  21981&            265&       A1V     &          5.513&          5.526&          41.35&           0.93\\
HD  34868&            300&       A0V     &          6.018&          6.024&           26.3&           0.93\\
HD  42525&            300&       A0V     &          5.763&          5.751&          34.76&           0.96\\
HD  57336&            400&       A0IV    &          7.188&          7.114&          9.544&           0.92\\
HD  73210&            729&       A5V     &          6.170&          6.165&          24.96&           1.01\\
HD  73666&            729&       A1V     &          6.524&          6.532&          18.11&           1.03\\
HD  73819&            729&       A6Vn    &          6.322&           6.28&          21.42&           0.96\\
HD  92845&            300&       A0V     &          5.519&          5.513&          47.71&           1.06**\\
HD 101452&            250&       A2      &          6.799&          6.819&          13.01&           0.96\\
HD 105805&            500&       A4Vn    &          5.626&            5.6&          38.96&           0.94\\
HD 116706&            500&       A3IV    &          5.507&          5.502&          42.03&           0.92\\
HD 128998&            250&       A1V     &          5.758&          5.756&          32.73&           0.91\\
HD 158485&            420&       A4V     &          6.153&          6.145&          24.30&           0.96\\
HD 163466&            310&       A2      &          6.300&          6.339&          19.48&           0.92\\
HD 163466&            310&       A2      &          6.300&          6.339&          20.03&           0.95\\
HD 172728&            210&       A0V     &          5.746&          5.753&          32.33&           0.89***\\
HD 172728&            210&       A0V     &          5.746&          5.753&           32.3&           0.89***\\
\enddata

\tablecomments{*Ratio of observed flux density to expected flux density from the photosphere. 
See Rieke et al. (2005) and Su et al. (2006) for further details. **Possible weak 24$\mu$m excess. ***Low value probably reflects a measurement error, possibly in the 2MASS K$_S$ magnitude.}

\end{deluxetable}

\begin{deluxetable}{lllllll}
\tablenum{C3}
\tablecolumns{7}
\tabletypesize{\scriptsize}
\tablecaption{ \label{Gstars} Solar Type 24$\mu$m Calibration Sample.}
\tablewidth{0pc}
\tablehead{\colhead{Name}  &   \colhead{age}  &  \colhead{Type } & 
\colhead{$SuperK_S$}  & \colhead{K$_S$}  &  \colhead{f(24$\mu$m)}  &  
\colhead{flux ratio*} \\
\colhead{}  &  \colhead{(Myr)}  &  \colhead{}  &   \colhead{(mag)}  &  \colhead{(mag)}  
&  \colhead{(mJy)}  &  \colhead{}  \\
}
\startdata
HD 008941   &      2200&    F8IV-V  &    5.353&       5.378&    49.2&    0.96\\
HD 019019   &      1700&    F8      &    5.563&        5.57&    42.4&    0.99\\
HD 027466   &      1500&    G5      &    6.262&       6.282&    21.5&    0.97\\
HD 037962   &      2400&    G4V     &    6.279&       6.268&    22.4&    1.00\\
HD 038949   &      1000&    G1V     &    6.419&       6.442&    19.6&    1.02\\
HD 064324   &      1700&    G0      &    6.234&       6.235&    25.1&    1.08\\
HD 066751   &      4200&    F8      &    5.061&       5.066&    66.1&    0.97\\
HD 092788   &      6800&    G5      &    5.724&       5.721&    36.2&    0.97\\
HD 098553   &      5800&    G2/G3V  &    6.072&       6.078&    26.4&    0.98\\
HD 100167   &      2300&    F8      &    5.826&       5.806&    34.4&    1.00\\
HD 101472   &      1100&    G0      &    6.125&       6.139&    26.2&    1.03\\
HD 101959   &      4700&    G0V     &    5.611&       5.609&    39.8&    0.96\\
HD 106252   &      5000&    G0V     &    5.940&       5.929&    32.4&    1.06\\
HD 108799   &      9300&    G1/G2V  &    4.857&       4.832&    84.9&    1.01\\
HD 108944   &      4500&    F8      &    6.009&       6.025&    30.3&    1.08\\
HD 112196   &      7300&    F8V     &    5.556&       5.553&    43.4&    1.00\\
HD 115043   &     11500&    G1Va    &    5.331&       5.334&    53.7&    1.01\\
HD 121504   &      7100&    G2V     &    6.135&       6.124&    27.4&    1.07\\
HD 122652   &      2500&    F8      &    5.863&       5.879&    34.8&    1.08\\
HD 141937   &      4500&    G2/G3V  &    5.771&       5.764&    35.0&    0.98\\
HD 150706   &      1400&    G0      &    5.572&       5.565&    45.5&    1.06\\
HD 153458   &      1400&    G0      &    6.449&       6.447&    19.1&    1.00\\
HD 159222   &      3800&    G5V     &    4.981&       4.998&    68.3&    0.94\\
HD 167389   &      2200&    F8      &    5.910&       5.918&    30.4&    0.98\\
HD 193017   &      1900&    F8      &    5.945&       5.966&    28.4&    0.95\\
HD 195034   &      4800&    G5      &    5.561&       5.579&    41.9&    0.99\\
HD 199598   &      2100&    G0V     &    5.462&       5.461&    48.0&    1.01\\
HD 204277   &      1100&    F8V     &    5.412&       5.449&    49.9&    1.04\\
HD 205905   &      1500&    G4IV-V  &    5.314&       5.321&    53.8&    1.00\\
HD 212291   &      2900&    G5      &    6.265&       6.269&    22.7&    1.01\\
HD 216275   &      4400&    G0      &    5.782&       5.794&    34.6&    0.99\\

\enddata

\tablecomments{*Ratio of observed flux density to expected flux density from the photosphere. 
See Rieke et al. (2005) and Su et al. (2006) for further details.}

\end{deluxetable}

\begin{deluxetable}{llllll}
\tablenum{D1}
\tablecolumns{6}
\tabletypesize{\scriptsize}
\tablecaption{ \label{Azero} Near Infrared Colors of A Stars in 2MASS Read 1 Mode.}
\tablewidth{0pc}
\tablehead{\colhead{Name}  &  \colhead{V} & 
\colhead{B - V}  & \colhead{V - K}  &  \colhead{J - K}  &  
\colhead{H - K} \\
}
\startdata HD 012468  &         6.52  &        0.003  &        -0.037 &        -0.05  &        0.016  \\
  HD 021379  &         6.29  &        -0.025 &        -0.021 &        -0.029 &         0.01  \\
  HD 022789  &         6.01  &        -0.027 &        0.002  &          0    &        0.057  \\
  HD 029526  &         5.66  &        0.002  &        0.011  &        0.014  &        0.019  \\
  HD 031069  &         6.06  &        -0.032 &        -0.047 &        -0.059 &        -0.004 \\
  HD 035505  &         5.64  &        -0.001 &        -0.009 &        -0.002 &        0.029  \\
  HD 035656  &         6.41  &        -0.023 &        -0.028 &        -0.044 &        -0.02  \\
  HD 036473  &         5.53  &        0.012  &        0.016  &        0.032  &        0.006  \\
  HD 041076  &         6.09  &        -0.035 &        0.002  &        0.004  &        0.035  \\
  HD 042301  &         5.49  &        -0.01  &        0.037  &        0.008  &        0.034  \\
  HD 042729  &         6.08  &        -0.021 &        0.051  &        0.004  &         0.05  \\
  HD 043583  &         6.59  &        -0.036 &        -0.031 &        -0.057 &        -0.009 \\
  HD 045137  &         6.51  &        -0.027 &          0    &        -0.011 &        -0.005 \\
  HD 045557  &         5.78  &        -0.001 &        0.026  &        0.001  &        0.049  \\
  HD 056386  &         6.19  &        -0.024 &        0.018  &        -0.042 &        0.016  \\
  HD 060629  &         6.64  &        -0.007 &        -0.015 &        -0.03  &        0.027  \\
  HD 070175  &          7    &        -0.01  &        -0.009 &        -0.001 &        0.035  \\
  HD 071043  &         5.89  &        0.018  &        0.017  &        -0.011 &        -0.002 \\
  HD 072337  &         5.51  &        -0.024 &        -0.04  &        -0.008 &        0.028  \\
  HD 076346  &         6.02  &        -0.023 &        -0.024 &        -0.004 &        0.003  \\
  HD 078955  &         6.53  &        0.001  &        0.041  &        0.018  &        0.062  \\
  HD 079108  &         6.14  &        -0.01  &        0.021  &        0.023  &        0.044  \\
  HD 080950  &         5.86  &        -0.024 &        -0.005 &        0.035  &        0.055  \\
  HD 086087  &         5.71  &        0.006  &        -0.027 &        -0.035 &        0.028  \\
  HD 096338  &         6.82  &        0.043  &        0.051  &        0.009  &        -0.007 \\
  HD 102981  &         6.62  &        -0.027 &        -0.021 &        -0.026 &         0.03  \\
  HD 104430  &         6.16  &        -0.004 &        -0.034 &        -0.01  &        -0.015 \\
  HD 107655  &         6.21  &        -0.012 &        0.043  &        -0.049 &        -0.029 \\
  HD 107947  &         6.61  &        -0.001 &        -0.008 &        -0.037 &        0.032  \\
  HD 113457  &         6.64  &        0.006  &          0    &        -0.032 &        0.007  \\
  HD 115527  &         6.87  &        -0.006 &        -0.007 &        -0.032 &        0.011  \\
  HD 117651  &         6.36  &        -0.014 &        -0.006 &        -0.044 &         0.03  \\
  HD 118214  &         5.6   &        -0.014 &        -0.035 &        -0.035 &        0.018  \\
  HD 121409  &         5.7   &        -0.032 &        0.022  &        -0.03  &         0.03  \\
  HD 124683  &         5.53  &        0.001  &        -0.022 &        -0.009 &        -0.004 \\
  HD 131951  &         5.9   &        -0.031 &        0.001  &        -0.048 &         0.02  \\
  HD 136831  &         6.29  &        -0.008 &        0.042  &        -0.031 &        0.008  \\
  HD 140729  &         6.15  &        0.002  &        0.019  &        -0.054 &        -0.016 \\
  HD 143488  &         6.99  &        0.006  &        0.023  &        -0.022 &        0.042  \\
  HD 145122  &         6.13  &        -0.003 &        0.046  &        -0.032 &        0.005  \\
  HD 145454  &         5.44  &        -0.019 &        0.009  &        -0.058 &        -0.003 \\
  HD 154972  &         6.24  &        -0.004 &         0.02  &        -0.011 &        0.014  \\
  HD 155379  &         6.52  &        -0.037 &          0    &        -0.007 &        -0.029 \\
  HD 172728  &         5.74  &        -0.045 &        -0.013 &        -0.028 &         0.02  \\
  HD 176425  &         6.21  &        0.001  &        -0.018 &        -0.024 &        0.064  \\
  HD 177406  &         5.95  &        -0.017 &        -0.014 &        -0.006 &        0.031  \\
  HD 178207  &         5.4   &        -0.014 &        -0.014 &        -0.046 &        0.032  \\
  HD 179933  &         6.26  &        0.008  &        0.001  &        -0.048 &        0.003  \\
  HD 182761  &         6.31  &        -0.014 &        0.023  &        -0.045 &        0.031  \\
  HD 182919  &         5.6   &        -0.006 &        -0.013 &        -0.038 &        0.014  \\
  HD 195549  &         6.35  &        0.001  &        0.007  &        -0.007 &        0.017  \\
  HD 205314  &         5.77  &        -0.042 &        -0.018 &        -0.055 &        0.025  \\
  HD 207636  &         6.45  &        -0.007 &        0.033  &        -0.015 &        0.026  \\
  HD 212643  &         6.29  &        -0.021 &        0.005  &        -0.028 &        -0.006 \\
  HD 219290  &         6.32  &        -0.005 &        -0.001 &        -0.024 &        -0.006 \\
  HD 219485  &         5.89  &        -0.01  &        0.018  &        -0.035 &        0.026  \\
  HD 223386  &         6.33  &        -0.007 &        -0.001 &        -0.03  &        -0.005 \\

\enddata
\end{deluxetable}

\begin{deluxetable}{lllll}
\tablenum{D2}
\tablecolumns{5}
\tabletypesize{\scriptsize}
\tablecaption{ \label{solarclone} Properties of Solar Analog Stars.}
\tablewidth{0pc}
\tablehead{\colhead{Name}  &  \colhead{T$_{eff}$} & 
\colhead{log(g)}  & \colhead{M/H}  &  \colhead{V - K$_S$}  \\
}
\startdata
Sun     &           5778&           4.44&              0&           1.57\\
HD 001562&           5756&           4.43&          -0.22&           1.54\\
HD 003821&           5785&           4.40&          -0.07&           1.50\\
HD 008262&           5861&           4.34&          -0.13&           1.57\\
HD 009986&           5749&           4.39&          -0.03&           1.55\\
HD 010086&           5659&           4.61&          -0.02&           1.62\\
HD 010145&           5673&           4.40&          -0.01&           1.64\\
HD 011131&           5819&           4.53&          -0.02&           1.57\\
HD 012846&           5667&           4.38&          -0.25&           1.60\\
HD 020619&           5666&           4.50&          -0.25&           1.58\\
HD 041330&           5858&           4.25&          -0.20&           1.52\\
HD 042618&           5714&           4.58&          -0.16&           1.55\\
HD 063433&           5691&           4.60&          -0.03&           1.64\\
HD 071148&           5756&           4.35&          -0.02&           1.49\\
HD 073350&           5754&           4.37&           0.04&           1.52\\
HD 075767&           5741&           4.42&          -0.08&           1.57\\
HD 088072&           5746&           4.31&           0.01&           1.55\\
HD 089269&           5674&           4.40&          -0.23&           1.65\\
HD 090508&           5779&           4.24&          -0.23&           1.55\\
HD 114174&           5750&           4.34&           0.07&           1.58\\
HD 139777&           5703&           4.55&          -0.05&           1.50\\
HD 142093&           5859&           4.55&          -0.15&           1.49\\
HD 146946&           5854&           4.21&          -0.21&           1.62\\
HD 147044&           5849&           4.29&          -0.03&           1.45\\
HD 159222&           5788&           4.25&           0.12&           1.52\\
HD 165401&           5798&           4.27&          -0.29&           1.55\\
HD 166435&           5741&           4.41&          -0.07&           1.52\\
HD 168009&           5801&           4.09&          -0.04&           1.54\\
HD 177082&           5733&           4.25&          -0.11&           1.60\\
HD 186427&           5753&           4.25&           0.06&           1.60\\
HD 187237&           5788&           4.61&           0.06&           1.53\\
HD 196850&           5792&           4.30&          -0.12&           1.53\\
HD 197076&           5842&           4.34&          -0.02&           1.51\\
HD 202108&           5665&           4.38&          -0.30&           1.57\\
HD 217813&           5861&           4.41&          -0.03&           1.50\\
HD 218739&           5788&           4.41&           0.06&           1.44\\
HD 224465&           5664&           4.22&          -0.12&           1.58\\

\enddata
\end{deluxetable}

\begin{deluxetable}{lll}
\tablenum{E1}
\tablecolumns{3}
\tabletypesize{\scriptsize}
\tablecaption{ \label{interpolation} Beginning and end wavelengths for isophotal interpolations}
\tablewidth{0pc}
\tablehead{\colhead{}  &  \colhead{Beginning}  &  \colhead{End} \\
}
\startdata
J & 1.221$\mu $m & 1.240$\mu $m \\
H & 1.6255$\mu $m & 1.660$\mu $m \\
K$_{S}$  & 2.134$\mu $m & 2.200$\mu $m \\
\enddata
\end{deluxetable}


\begin{thebibliography}{}

\bibitem[]{} 2MASS Web site: \underline{http://www.ipac.caltech.edu/2mass/}

\bibitem[]{} Absil, O., et al. 2006, A\&A, 452, 237

\bibitem[]{} Anders, E., Grevesse, N., 1989, Geochimica et Cosmochimica Acta 53, 197

\bibitem[]{} Aufdenberg, J. P., et al. 2006, ApJ, 645, 664

\bibitem[]{} Aumann, H. H., et al. 1984, ApJL, 278, 23

\bibitem[]{} Ayres, T. R., Plymate, C., \& Keller, C. U. 2006, ApJS, 165, 618

\bibitem[]{} Becklin, E. E., Hansen, O., Kieffer, H., and Neugebauer, G. 1973, AJ, 78, 1063 

\bibitem[]{} Beichman, C. A., Neugebauer, G., Habing, H. J., Clegg, P. E., and Chester, T. J. 1988, \textit{IRAS Explanatory Supplement}, NASA: Washington, D. C. Section VI. C.2.a

\bibitem[]{} Bell, K.L., \& Berrington, K.A., 1987, J. Phys. B20, 801 

\bibitem[]{} Bessell, M. S., \& Brett, J. M. 1988, PASP, 100, 1134

\bibitem[]{} Bessell, M. S., Castelli, F., and Plez, B. 1998, A\&A, 333, 231 


\bibitem[]{} Bessell, M. S. 2005, ARAA, 43, 293 

\bibitem[]{} Blackwell, D. E., Leggett, S. K., Petford, A. D., Mountain, C. M., \& Selby, M. J. 1983, MNRAS, 205, 897 

\bibitem[]{} Bohlin, R. C. 2007, in {\it The Future of Photometric Systems}," PASP, 364, 315

\bibitem[]{} Bohlin, R. C., Dickinson, M. E., and Calzetti, D. 2001, AJ, 122, 2118

\bibitem[]{} Bohlin, R. C., and Gilliland, R. L. 2004, AJ, 127, 3508

\bibitem[]{} Booth, A. J., Selby, M. J., Blackwell, D. E., Petford, A. D., \& Arribas, S. 1989, A\&A, 218, 167 

\bibitem[]{} Bouchet, P., Schmider, F. X., and Manfroid, J. 1991, A\&AS, 91, 409 

\bibitem[]{} Campins, H., Rieke, G. H., \& Lebofsky, M. J. 1985, AJ, 90. 896 

\bibitem[]{} Carpenter, J. M. 2001, AJ, 121, 2851

\bibitem[]{} Casagrande, L., Portinari, L., \& Flynn, C. 2006, MNRAS, 373, 13

\bibitem[]{} Carter, B. S. 1990, MNRAS, 242, 1 

\bibitem[]{} Cayrel de Strobel, G., 1996, A\&A Rev. 7, 243 

\bibitem[]{} Ciardi, D. R., van Belle, G. T., Akeson, R. L., Thompson, R. R., Lada, E. A., and Howell, S. B. 2001, ApJ, 559, 1147 

\bibitem[]{} Cohen, M., Walker, R. G., Barlow, M. J., and Deacon, J. R. 1992, AJ, 104, 1650 

\bibitem[]{} Cohen, M., Wheaton, W. A., \& Megeath, S. T. 2003, AJ, 126, 1090

\bibitem[]{} Colina, L., Bohlin, R. C., and Castelli, F. 1996, AJ 112, 307 

\bibitem[]{} Cutri, R. M. et al. 2003, \textit{The IRSA 2MASS All-Sky Point Source Catalog}

\bibitem[]{} Decin, L., 2000, PhD Thesis, University of Leuven (Belgium)

\bibitem[]{} Decin, L., Vandenbussche, B., Waelkens, C., Eriksson, K., Gustafsson, B., Plez, B., Sauval, A.J., and Hinkle, K., 2003, A\&A 400, 679 

\bibitem[]{} Elias, J. H., Frogel, J. A., \& Humphreys, R. M. 1985, ApJS, 57, 91

\bibitem[]{} Elias, J. H., Frogel, J. A., Matthews. K., and Neugebauer, G. 1982, AJ, 87, 1092

\bibitem[]{} Engelke, C. W. 1992, AJ, 104, 1248

\bibitem[]{} Engelbracht, C., et al. 2007, PASP, 119, 994

\bibitem[]{} Fazio, G. G., et al. 2004, ApJS, 154, 39


\bibitem[]{} Fontenla, J. M., Avrett, E., Thuillier, G., and Harder, J. 2006, ApJ, 639, 441

\bibitem[]{} Golay, M. 1974, \textit{Introduction to Stellar Photometry,} (D. Reidel: Dordrecht), p. 40


\bibitem[]{} Gray, R. O. 1998, PASP, 116, 482

\bibitem[]{} Gray, R. O. 2007, \underline {http://stellar.phys.appstate.edu/}

\bibitem[]{} Gulliver, A. F., Hill, G., \& Adelman, S. J. 1994, ApJL, 429, L81

\bibitem[]{} Hammersley, P. L., Jourdain de Muizon, M., Kessler,M. F., 
Bouchet, P., Joseph, R. D., Habing, H. J., Salama, A., \& Metcalfe, L. 
1998, A\&AS, 128, 307

\bibitem[]{} Harris, M.J., Lambert, D.L., \& Goldman, A., 1987, MNRAS224, 237 

\bibitem[]{} Holmberg, J., Flynn, C., and Portinari, L. 2006, MNRAS, 367, 449

\bibitem[]{} Holweger, H. and M\"uller E.A., 1974, Solar Physics 39, 19

\bibitem[]{} Johnson, H. L. 1965a, Comm. Lunar Plan. Lab., 3, 73

\bibitem[]{} Johnson, H. L. 1965b, ApJ, 141, 923

\bibitem[]{} Johnson, H. L., and Morgan, W. W. 1953, ApJ, 117, 313

\bibitem[]{} Johnson, H. L., Iriarte, B., Mitchell, R. I., and Wisniewski, W. Z. 1966, Comm. Lunar Plan. Lab., 4, 99

\bibitem[]{} Johnson, H. L., MacArthur, J. W., \& Mitchell, R. I. 1968, ApJ, 
152, 465

\bibitem[]{} Kidger, M. R., \& Mart\'in-Luis, F. 2003, AJ, 125, 3311

\bibitem[]{} Knude, J., \& Hog, E. 1998, A\&A, 338, 897

\bibitem[]{} Koornneef, J. 1983, A\&AS, 51, 489

\bibitem[]{} Kurucz, R. L. 2005, \underline{http://kurucz.harvard.edu}

\bibitem[]{} Labs, D. \& Neckel, H. 1968, Zeitschrift f\'ur Astrophysik, 69, 1

\bibitem[]{} Labs, D. \& Neckel, H. 1970, Solar Physics, 15, 79

\bibitem[]{} Lee, T. A. 1968, ApJ, 152, 913

\bibitem[]{} Low, F. J., and Rieke, G. H. 1974, in \textit{Methods of Experimental 
Physics,} Vol. 12, Part A, edited by N. Carleton (Academic: New York) p. 456

\bibitem[]{} Maltby, P., Avrett, E. H., Carlsson, M., Kjeldseth-Moe, O., Kurucz, R. L., \& Loesser, R. 1986, ApJ, 306, 284

\bibitem[]{} M\'egessier, C. 1995, A\&A, 296, 771

\bibitem[]{} MIPS Data Handbook: \underline{http://ssc.spitzer.caltech.edu/mips/dh}

\bibitem[]{} Mountain, C. M., Selby, M. J., Leggett, S. K., Blackwell, D. E., \& Petford, A. D. 1985, A\&A, 151, 399

\bibitem[]{} NASA/IPAC Infrared Science Archive 2007, http://irsa.ipac.caltech.edu/data/SPITZER/FEPS/



\bibitem[]{} Perryman, M. A. C. et al. 1997, A\&A, 323, L49

\bibitem[]{} Plez, B., Smith, V.V., Lambert, D.L., 1993, ApJ 418, 812 

\bibitem[]{} Price, S. D., Paxson, C., Engelke, C., \& Murdock, T. L. 2004, AJ, 128, 889

\bibitem[]{} Price, S. D. 2004, Space Science Reviews, 113, 409

\bibitem[]{} Reach, W. T. et al. 2005, PASP, 117, 978

\bibitem[]{} Rieke, G. H., Lebofsky, M. J., \& Low, F. J. 1985, AJ, 90, 900

\bibitem[]{} Rieke, G. H., \& Lebofsky, M. J. 1985, ApJ, 288, 618

\bibitem[]{} Rieke, G. H. et al. 2004, ApJS, 154, 25

\bibitem[]{} Rieke, G. H. et al. 2005, ApJ, 620, 1010

\bibitem[]{} Saiedy, F. 1960, MNRAS, 121, 483

\bibitem[]{} Saeidy, F., and Goody, R. M. 1959, MNRAS, 119, 213

\bibitem[]{} Selby, M. J., Mountain, C. M., Blackwell, D. E., Petford, A. D., \& Leggett, S. K. 1983, MNRAS, 203, 795

\bibitem[]{} Soubiran, C. \& Triaud, A. 2004, A\&A, 418, 1089

\bibitem[]{} Spitzer Science Center 2007, http://ssc.spitzer.caltech.edu/postbcd/spice.html/


\bibitem[]{} Su, K. Y. L., et al. 2006, ApJ, 653, 675

\bibitem[]{} Thuillier, G., Hers\'e, M., Labs, D., Foujols, T., Peetermans, W., Gillotay, 
D., Simon, P. C., \& Mandel, H. 2003, Solar Phys., 214, 1

\bibitem[]{} Tokunaga, A. T. 2000, in \textit{Allen's Astrophysical Quantities,} 4$^{th}$ 
edition, ed. A. N. Cox, Springer-Verlag: NY, p. 143

\bibitem[]{} Tokunaga, A. T., \& Vacca, W. D. 2005, PASP, 117, 421



\bibitem[]{} Vernazza, J. E., Avrett, E. H., \& Loeser, R. 1976, ApJS, 30, 1

\bibitem[]{} Walker, R. G. 1969, Phil. Trans. Roy. Soc. A, 264, 209 

\bibitem[]{} Wallace, L., and Livingston, W. 2003, National Solar Obsevatory Technical Report \#03-001

\bibitem[]{} Wright, J. T., Marcy, G. W., Butler, R. P., and Vogt, S. S. 2004, ApJS, 152, 261

\end{thebibliography}
\end{document}